\documentclass[reprint, amsmath, amssymb, aps, nofootinbib]{revtex4-2}
\usepackage[pdftex]{graphicx}
\usepackage{bm}
\usepackage{here}

\begin{document}
\title{Numerical analysis of a baryon and its dilatation modes in holographic QCD}
\author{Keiichiro~Hori}
\author{Hideo~Suganuma}
\affiliation{
Department of Physics, 
Kyoto University, \\ 
Kitashirakawaoiwake, Sakyo, Kyoto 606-8502, Japan}
\author{Hiroki~Kanda}
\affiliation{Yukawa Institute for Theoretical Physics (YITP), 
Kyoto University, 
Kitashirakawaoiwake, Sakyo, Kyoto 606-8502, Japan}
\date{\today} 
\begin{abstract}
We investigate a baryon and its dilatation modes 
in holographic QCD based on the Sakai-Sugimoto model, 
which is expressed as 
a 1+4 dimensional U($N_f$) gauge theory 
in the flavor space.
For spatially rotational symmetric systems, 
we apply a generalized version of the Witten Ansatz, 
and reduce 1+4 dimensional holographic QCD into a 1+2 dimensional Abelian Higgs theory in a curved space.
In the reduced theory, the holographic baryon is described  
as a two-dimensional topological object of an Abrikosov vortex.
We numerically calculate the baryon solution of holographic QCD 
using a fine and large lattice with 
spacing of 0.04~fm and size of 10~fm. 
Using the relation between the baryon size and 
the zero-point location of the Higgs field 
in the description with the Witten Ansatz,
we investigate a various-size baryon through this vortex description. 
As time-dependent size-oscillation modes 
(dilatation modes) of a baryon, we numerically 
obtain the lowest excitation energy of 577~MeV 
and deduce the dilatational excitation of a nucleon 
to be the Roper resonance N$^*$(1440). 

\end{abstract}

\maketitle

\section{Introduction}

Quantum chromodynamics (QCD) is established as 
the fundamental theory of strong interaction and 
characterized by ${\rm SU}(N_c)$ gauge symmetry and 
global ${\rm SU}(N_f)_L\times {\rm SU}(N_f)_R$ chiral symmetry. 
Owing to asymptotic freedom of QCD, 
high-energy hadron reactions can be analyzed using perturbative QCD.
In low energy regions, however, the QCD coupling becomes strong, 
and the perturbative method is no more applicable.
Therefore, for theoretical analyses of hadrons based on QCD, 
some nonperturbative methods are necessary such as lattice QCD.
Based on gauge/gravity duality \cite{M98} for D branes \cite{P95}
in superstring theory, holographic QCD is an interesting new tool 
to analyze the nonperturbative properties of QCD \cite{W98, KMMW04, SS05}.

Around 1970, the string theory was originally proposed by Nambu, Goto, and Polyakov for the description of hadrons \cite{Nambu,Goto,Polyakov81}. 
After establishment of QCD, the string theory was not used for the main research of hadrons. Instead, this framework was reformulated as the superstring theory in 1980s \cite{Green-Schwarz-Witten}
and has been studied as a plausible candidate of a grand unified theory including quantum gravity, and many studies have been constantly conducted to date. 

The superstring theory is formulated in 10 dimensional space-time 
and includes D branes, on which open string endpoints exist \cite{P95}. 
As a remarkable discovery by Polchinski, 
on the surface of $N$ overlapped D branes, 
U($N$) gauge symmetry emerges, 
and the $N$ D-branes system leads to U($N$) gauge theory \cite{P95}. 
In fact, on the $N$ D-branes, 
gauge fields $A_{ab}$ appear from the open string 
linking two D-branes labeled with $a$ and $b$ ($a,b =1,2, ..., N$).
On the other hand, around the D brane, 
a higher-dimensional supergravity theory is formed 
since the brane has mass, and multiple branes can be gravitational sources. 
Thus, there are two different theories relating to the D-brane,
and the gauge theory {\it on} the D-brane and 
the higher-dimensional gravity theory {\it around} the D-brane 
are conjectured to be equivalent \cite{M98}, 
which is called AdS/CFT correspondence or gauge/gravity duality. 
This remarkable equivalence between gauge and gravity theories 
was first proposed by Maldacena 
from a detailed analysis of four-dimensional ${\cal N}=4$ 
supersymmetric (SUSY) SU($N$) Yang-Mills theory 
and five-dimensional Anti-de Sitter (AdS) supergravity theory \cite{M98},
using large $N$ argument. 
This equivalence was first called as AdS/CFT correspondence 
and is generalized as gauge/gravity duality in more general concept. 
In this correspondence, 
a strong-coupling gauge theory can be described with 
a weak-coupling gravity theory \cite{M98,W98,KMMW04,SS05}.

With this correspondence, many researches have been performed, 
and one of the main category is to analyze strong-coupling QCD 
with a higher-dimensional classical gravity theory, 
which is called holographic QCD \cite{W98, Karch-Katz, EKSS, KMMW04, SS05}. 
In 1998, Witten formulated a non-SUSY version of holographic QCD 
for a four-dimensional pure-gluon Yang-Mills theory 
using $S^1$-compactified 
$N_c$ D4 branes, where periodic and antiperiodic boundary conditions are imposed on boson and fermion fields, respectively \cite{W98}.
In 2005, Sakai and Sugimoto proposed holographic QCD for 
four-dimensional full QCD, 
including massless chiral quarks, by adding the $N_f$ pair of 
${\rm D8}$ and $\overline{\rm D8}$ branes to $S^1$-compactified ${\rm D4}$ branes \cite{SS05}.
In fact, the D4/D8/$\overline{\rm D8}$ multi-D-brane system 
has SU($N_c$) gauge symmetry and 
${\rm SU}(N_f)_L\times {\rm SU}(N_f)_R$ chiral symmetry  
and leads to massless QCD in infrared scale.
Here, quarks and gluons appear from the massless modes 
of 4-8 and 4-4 strings, respectively.
In the large $N_c$ limit, 
$N_c$ D4 branes are dominant as a gravity source and 
are converted into a gravitational field 
via the AdS/CFT correspondence or the gauge/gravity duality. 
The 't~Hooft coupling $\lambda \equiv N_c g^2$ 
is a control parameter of the gauge side, 
and strong-coupling QCD with a large $\lambda$ 
corresponds to a weakly interacting gravitational theory. 
Then, this system is described by the D8 brane 
in the presence of a background gravitational field 
originating from the D4 brane. 
Nonperturbative properties of 
QCD can be analyzed with a classical gravity theory 

In the leading order of 1/$N_c$ expansion, 
the effective theory of the D8 brane in the D4-brane-induced 
background gravity is expressed with the Dirac-Born-Infeld (DBI) action and the Chern-Simons (CS) term. 
The leading order of $1/\lambda$ expansion is the DBI action, 
and the next leading of $1/\lambda$ is the CS term. 
Expanding the DBI action with $1/\lambda$, 
the leading order becomes 
a five-dimensional Yang-Mills theory 
in flavor space in a curved space, 
where a background gravity appears 
only in the fifth dimension. 
The CS term is a topological term 
responsible for anomalies in QCD. 
In holographic QCD, 
the five-dimensional holographic field is 
decomposed into four-dimensional meson fields, 
and this theory is successful in the mesons sector, 
that is, it describes low-lying meson masses, 
inter-meson couplings, and phenomenological laws 
in hadron physics \cite{SS05}. 
Regarding baryonic degrees of freedom, 
baryons do not appear explicitly in the holographic action. 
In fact, the baryon does not appear as a system component 
in holographic QCD. 

The absence of baryons in the effective action is 
a general result due to the large $N_c$ argument, 
because QCD is reduced into a weak-coupling theory 
of mesons and glueballs in the large $N_c$ limit \cite{W79}, 
where the baryon mass increases as ${\mathcal O}(N_c)$ and 
the baryon do not appear in the effective action. 
In the large $N_c$ argument, 
the baryon is considered to appear as a soliton of mesons 
such as the Skyrmion in the Skyrme model \cite{S61,W79,ZB96}. 

Here, we briefly mention a historical overview of the Skyrme model.
The Skyrme model is a low-energy effective theory 
in hadron physics, first proposed by Skyrme in 1961 \cite{S61}. 
After the quark model was proposed in 1964 \cite{GZ64}, 
main researches of strong interaction and hadrons were shifted to 
the quark theory, which was eventually developed into QCD. 
In 1979, Witten revived the Skyrme soliton picture of baryons 
from a large $N_c$ viewpoint of QCD \cite{W79}, 
although the direct relation between the Skyrme model and 
QCD has not been clear. 
In 2005, Sakai and Sugimoto showed 
a theoretical explicit connection between 
massless QCD and the Skyrme Lagrangian, 
which is derived as the pion sector in holographic QCD, 
using the gauge/gravity duality for a D-brane system \cite{SS05}. 
(Actually, the Sakai-Sugimoto model reduces into the Skyrme model, 
when massive mesons are dropped off, leaving only light pions.) 

Now, let us concentrate on baryons in holographic QCD. 
Also for holographic QCD, 
which is derived with large $N_c$ and is 
described with only meson fields, 
the baryon is described as a chiral soliton of mesons, 
that is, a topological object 
like a brane-induces Skyrmion \cite{NSK07, NSK09} or 
an instanton-like object \cite{W98b, HSSY07}. 
In fact, the baryon appears as 
an spatially extended object in holographic QCD.

To summarizes the above, 
holographic QCD is an analytical nonperturbative method for QCD 
and has direct connection with QCD, 
as a clearly strong advanced point. 
In holographic QCD, 
while mesons appear in the action and can be directly treated, 
the baryon is described as an extended soliton of mesons.
Therefore, compared with the meson sector, 
the baryon sector is more difficult and 
has not been well studied in holographic QCD, 
in spite of several pioneering studies on the holographic baryon 
\cite{NSK07,HSSY07,HRYY07,HSS08,CI12,BS14,RSR14}. 
In addition, this baryonic soliton allows 
spatial dilatation modes as its excitation 
peculiar to the spatially-extended object, 
and thus we focus on this dilatation mode of 
the holographic baryon in the latter of our study. 

In this study, 
we investigate a single baryon and its dilatation modes 
in holographic QCD, adopting the Sakai-Sugimoto model 
formulated as a 1+4 dimensional U($N_f$) gauge theory 
in the flavor space.
For spatially rotational symmetric systems, 
we apply a generalized Witten Ansatz and reduce  
1+4 dimensional holographic QCD into 
a 1+2 dimensional Abelian Higgs theory, 
where the holographic baryon is expressed 
as an Abrikosov vortex.
We numerically calculate the baryon solution of holographic QCD 
using a fine and large lattice, 
keeping background gravity from the $N_c$ D4-brane. 

In addition, we investigate a various-size baryon 
and the dilatation mode (time-dependent size-oscillation) 
of a single baryon, 
using the relation between the baryon size and 
the zero-point location of the Higgs field 
in the reduced Abelian Higgs theory. 

This size oscillation is physically considered 
as a collective motion and 
is difficult to be described in the quark model.
Instead, the size oscillation mode has been studied 
in the Skyrmion research, 
and its lowest excitation mode is identified 
as the Roper resonance N$^*$(1440) \cite{Roper,HS84,HH84,ZMK84,WE84,LZB84,BN84}. 

The Roper resonance is the first excited state 
of the nucleon N(940) with the positive parity and its energy being 1440 MeV. 
In the quark model, based on the single-particle picture, 
the first excited-state baryon is to have negative parity, 
and it contradicts the experimental data. 
In lattice QCD, the numerical results with overlap fermion 
well reproduce the Roper resonance in terms of the excitation energy 
and the positive parity \cite{MCDDHLLZ,LHKLMWZ}. 
Here, lattice QCD is a powerful tool for the quantitative analysis of hadrons, 
but it is difficult to get state information of hadrons like the wave function 
due to path integral formalism, where all the states are integrated out. 
As an alternative method, 
the Skyrme model seems to succeed to reproduce 
the mass and parity of N$^*$(1440) as the first excited state of N(940). 
In this chiral soliton picture, this first excited state 
is described as a dilatation or breathing mode of the ground-state soliton \cite{BN84,HS84,HH84,ZMK84,WE84,LZB84}. 

Therefore, we here investigate the dilatation mode of the baryon 
in holographic QCD and finally compare it with the Roper resonance N$^*$(1440) 
in terms of its mass and parity. 
The dilatation mode of baryon can be described also in holographic QCD. 
Note again that the Sakai-Sugimoto model reduces into the Skyrme model, 
when massive mesons except for pions are dropped off.
In fact, the holographic baryon appears as a soliton \cite{NSK07,HSSY07}, 
i.e., a spatially extended object, and therefore has dilatation mode. 
Note, however, that holographic dilatation 
is four-dimensional spatial oscillation including 
the extra spatial dimension rather than ordinary three-dimensional one. 

The organization of this paper is as follows.
In Sec.~II, 
we briefly review the Sakai-Sugimoto model as 
typical holographic QCD.
In Sec.~III, 
we apply the Witten Ansatz in holographic QCD.
Owing to the Witten Ansatz, the 1+4 dimensional Yang-Mills theory 
reduces into a 1+2 dimensional Abelian Higgs theory. 
In Sec.~IV, using the Witten Ansatz, 
we present the vortex description of baryons in holographic QCD, 
and we numerically obtain the ground-state solution of the holographic baryon
using a fine and large-volume lattice.
In Sec.~V, we numerically analyze size dependence 
of the holographic baryon. 
In Sec.~VI, we investigate time-dependent dilatational modes 
of a single baryon in holographic QCD. 
Section~VII is devoted for summary and conclusion.

\section{Holographic QCD action in the Sakai-Sugimoto model}

In this section, as a starting point, 
we introduce the Sakai-Sugimoto model, 
one of the most successful holographic QCD \cite{SS05}. 
In the Sakai-Sugimoto model, 
four-dimensional massless QCD is constructed 
using the D4/D8/$\overline{\rm D8}$ multi-D-brane system, 
which comprises spatially $S^1$-compactified $N_c$ D4 branes 
attached with $N_f$ D8-$\overline{\rm D8}$ pairs.
Here, $N_c$ means the color number, and 
$N_f$ the light flavor number. 
This compactification breaks SUSY due to the (anti)periodic conditions for bosons(fermions), 
as was demonstrated for a D4 brane system by Witten \cite{W98}.
The compactification radius is $M_{\rm KK}^{-1}$, 
and this model parameter physically corresponds to 
a UV cutoff in holographic QCD.
This system is infrared equivalent to massless QCD, 
where chiral symmetry exists \cite{SS05}.
Using AdS/CFT correspondence (gauge/gravity duality), 
the $N_c$ D4 branes are transformed into a gravitational source, 
and the system becomes $N_f$ D8 branes in the D4 gravity background, 
which leads to the DBI and CS action at the leading order of $1/N_c$ 
within the probe approximation. 
In terms of $1/\lambda$ expansion, 
the DBI action includes its leading order, and the CS action is sub-leading.

From the multi-D-brane system which is infrared equivalent to massless QCD, 
the DBI action becomes 1+4 dimensional Yang-Mills theory on the flavor space of 
U($N_f$) $\simeq {\rm SU}(N_f) \times$ U(1) 
at the leading order of $1/\lambda$ expansion \cite{SS05}: 
\begin{eqnarray}
&&S_{\rm 5YM} = S_{\rm 5YM}^{{\rm SU}(N_f)} +S_{\rm 5YM}^{\rm U(1)}
\nonumber \\
&=& -\kappa \int d^4x dw \ \textrm{tr} 
\left[ \frac{1}{2}h(w)F_{\mu\nu} F^{\mu\nu} + k(w)F_{\mu w} F^{\mu w} \right]
\nonumber \\
&&-\frac{\kappa}{2} \int d^4x dw \ \left(\frac{1}{2}h(w) \hat F_{\mu\nu} \hat F^{\mu\nu} 
+ k(w) \hat F_{\mu w} \hat F^{\mu w} \right).~~~~~
\label{eq:5YM}
\end{eqnarray}
In this paper, we use $w$ for the extra fifth-coordinate in holographic QCD.  
$\hat{A}$ denotes U(1) gauge field and $\hat{F}$ U(1) field strength \cite{SH20}. 

For $M, N = t, x, y, z, w$, the field strengths are given by 
\begin{eqnarray}
F_{MN} &\equiv& \partial_M A_N-\partial_N A_M+i[A_M, A_N], 
\nonumber \\
\hat F_{MN} &\equiv& \partial_M \hat A_N-\partial_N \hat A_M, 
\end{eqnarray}
with the five-dimensional SU($N_f$) gauge field $A^M(x^\mu,w)$ 
and U(1) gauge field $\hat A^M(x^\mu,w)$, respectively. 
Throughout this paper, we take the $M_{\rm KK}=1$ unit together with the natural unit, 
and $\kappa$ is written as $\kappa = \frac{\lambda N_c}{216\pi^3}$ in this unit.
Note that, as a relic of $N_c$ D4-branes, there appear background gravity $k(w)$ and $h(w)$ 
depending on the extra fifth-coordinate $w$, 
\begin{eqnarray}
k(w) \equiv 1+w^2, \ \ h(w) \equiv k(w)^{-1/3}, 
\end{eqnarray}
in the $M_{\rm KK}=1$ unit. 
In Eq.~(\ref{eq:5YM}) at the leading order of $1/\lambda$ expansion, 
SU($N_f$) variables $A$ and U(1) variables $\hat A$ are completely separated, 
and hence we have divided $S_{\rm 5YM}$ into the SU($N_f$) sector $S_{\rm 5YM}^{{\rm SU}(N_f)}$ 
and the U(1) sector $S_{\rm 5YM}^{{\rm U}(1)}$. 

The $1/N_c$-leading holographic QCD 
also has the CS term \cite{SS05,HSSY07}
as the next leading order of $1/\lambda$.
The CS term is a topological term responsible 
for anomalies in QCD, and its explicit form is 
\begin{eqnarray}
\label{eq:CS}
S_{\rm CS} &=& \frac{N_c}{24\pi^2} \int \omega_5({\cal A}) \nonumber \\
&=& \frac{N_c}{24\pi^2} \int {\rm tr} \left( {\cal A}{\cal F}^2 - \frac{i}{2}{\cal A}^3{\cal F} - \frac{1}{10}{\cal A}^5 \right), 
\end{eqnarray}
where ${\cal A} = A + \frac{1}{\sqrt{2N_f}}\hat{A}$
denotes the U($N_f$) gauge field \cite{SS05}.
Note that SU($N_f$) variables $A$ and U(1) variables $\hat A$ are dynamically mixed 
in the CS term $S_{\rm CS}$ in Eq.~(\ref{eq:CS}).

In this paper, to analyze baryons in holographic QCD, 
we consider both Yang-Mills and CS parts for the case of $N_f=2$. 

\section{Witten Ansatz in holographic QCD}

Holographic QCD is formulated to be a 1+4 dimensional 
U($N_f$) non-Abelian gauge theory with a gravitational background $h(w)$ and $k(w)$, 
which would be fairly difficult to analyze. 
To avoid the difficulty and to proceed analytic calculations, 
most previous works \cite{HSSY07,HRYY07,HSS08}
were forced to take a flat background $h(w)=k(w)=1$ 
and to use the simple 't~Hooft instanton solution 
\cite{BPST, tH76} in the flat space, 
although $h(w)$ and $k(w)$ are the trace of D4-branes 
and are to be relevant ingredients. 

To deal with holographic QCD for $N_f=2$ 
without reduction the gravitational backgrounds $h(w)$ and $k(w)$, 
we adopt the Witten Ansatz \cite{W77} in this paper. 
The Witten Ansatz is generally applicable for spatially-rotational symmetric system 
in the SU(2) Yang-Mills theory. 
Applying this to holographic QCD, the 1+4 dimensional non-Abelian theory 
transforms to a 1+2 dimensional Abelian Higgs theory. 
Accordingly, relevant topological objects are changed 
from instantons to vortices, as will be shown in Sec.~IV. 
In this section, we generalize the Witten Anstaz 
to be applicable to holographic QCD.

\subsection{Witten Ansatz for Euclidean Yang-Mills theory}

In this subsection, we briefly review the original Witten Ansatz \cite{W77}
applied for the Euclidean four-dimensional SU(2) Yang-Mills theory, 
which is formulated on three spatial coordinates $(x,y,z)$ and Euclidean time $t$.
For spatially-rotational symmetric systems, the Witten Ansatz can be applied as
\begin{eqnarray}
A_i^a(x,y,z,t) &=& \frac{\phi_2(r,t)+1}{r}\epsilon_{iak}\hat{x}_k + \frac{\phi_1(r,t)}{r} \hat{\delta}_{ia} \nonumber \\
& & + a_r(r,t) \hat{x}_i\hat{x}_a, \label{eqn:Original Witten Ansatz 1} \\
A_t^a(x,y,z,t) &=& a_t (r,t) \hat{x}^a 
\label{eqn:Original Witten Ansatz 2}
\end{eqnarray}
with $r \equiv (x_i x_i)^{1/2}$, $\hat x_i \equiv x_i/r$
and $\hat \delta_{ij} \equiv \delta_{ij}-\hat x_i \hat x_j$.

Using the Witten Ansatz, 
the four-dimensional SU(2) Yang-Mills theory is reduced into 
a two-dimensional Abelian Higgs theory as 
\begin{eqnarray}
S_{\rm YM}^{\rm SU(2)} &=& \int dt~d^3x\ \frac{1}{2} {\rm tr} F_{\mu\nu} F^{\mu\nu} \nonumber \\
&=& 4\pi \int^\infty_{-\infty} dt \int^\infty_0 dr \biggl[ |D_0\phi|^2 + |D_1\phi|^2 \nonumber \\
& & + \frac{1}{2r^2}(1-|\phi|^2)^2 + \frac{r^2}{2} f_{01}^2 \biggr],
\label{eq:YM->AH}
\end{eqnarray}
where the complex Higgs field $\phi(t,r)$, 
Abelian gauge field $a_\mu(t,r)$, its covariant derivative $D_\mu$, 
and field strength $f_{\mu\nu}$ in the Abelian Higgs theory are 
\begin{eqnarray}
\phi &\equiv& \phi_1+i\phi_2 \in {\bf C}, \
a_\mu \equiv (a_0,a_1), \cr 
D_\mu &\equiv& \partial_\mu - i a_\mu, \
f_{01} \equiv \partial_0 a_1-\partial_1 a_0. \ \
\end{eqnarray} 
Here, we have used $(0,1)=(t,r)$ for the index of 
two dimensional coordinates. 

\subsection{Generalized Witten Ansatz for 
SU(2)$_f$ sector in holographic QCD}

In this subsection, we generalize the Witten Ansatz 
to be applicable for holographic QCD.

The SU(2)$_f$ sector in holographic QCD is expressed as a 1+4 dimensional Yang-Mills theory with gravitational backgrounds $h(w)$ and $k(w)$. 
Holographic QCD already includes
four-dimensional Euclidean spatial coordinates ($x$, $y$, $z$, $w$) including the extra fifth-coordinate $w$, 
and instantons can be naturally introduced in holographic QCD without necessity of 
the Euclidean process or the Wick rotation. 

Describing the SU(2)$_f$ gauge field $A$ 
with the Pauli matrix $\tau^a$ as  
$A=A^a\frac{\tau^a}{2} \in {\rm su}(2)_f$ 
in holographic QCD, 
we take a generalized version of the Witten Ansatz 
for $(x,y,z)$-spatially rotational symmetric systems 
\cite{BS14,RSR14,SH20}, 
\begin{eqnarray}
A_0^a(t,x,y,z,w) &=& a_0 (t,r,w) \hat{x}^a, 
\label{eqn:Witten Ansatz 1} \\
A_i^a(t,x,y,z,w) &=& \frac{\phi_2(t,r,w)+1}{r}\epsilon_{iak}\hat{x}^k \cr
&+& \frac{\phi_1(t,r,w)}{r} \hat{\delta}_{ia} + a_r(t,r,w) \hat{x}_i\hat{x}_a,~~~ \label{eqn:Witten Ansatz 2} \\
A_w^a(t,x,y,z,w) &=& a_w (t,r,w) \hat{x}^a, 
\label{eqn:Witten Ansatz 3}
\end{eqnarray}
with $r \equiv (x_i x_i)^{1/2}$, $\hat x_i \equiv x_i/r$
and $\hat \delta_{ij} \equiv \delta_{ij}-\hat x_i \hat x_j$.
For 1+4 dimensional holographic QCD, we have extended 
the Witten Ansatz for $A^a_0$ component, 
considering the $(x,y,z)$-rotational symmetry. 
Note that this Ansatz is a general form 
when $(x,y,z)$-rotational symmetry is 
imposed on gauge fields. 

In the Witten Ansatz, 
the holographic field strength,  $F_{ij}$, $F_{0i}$, $F_{wi}$ and $F_{0w}$ 
are expressed as 
\begin{eqnarray}
\frac{1}{2}\epsilon_{ijk}F_{jk}^a &=& (\partial_1\phi_2 - a_1\phi_1)\frac{\hat{\delta}_{ai}}{r} + (1-\phi_1^2-\phi_2^2)\frac{\hat{x}_a\hat{x}_i}{r^2} \cr
 & & + (\partial_1\phi_1+a_1\phi_2)\frac{1}{r}\epsilon_{aik}\hat{x}_k, 
\\
F_{0i}^a &=& (\partial_0\phi_1+a_0\phi_2)\frac{\hat{\delta}_{ai}}{r} +  (\partial_0a_1-\partial_1a_0) {\hat{x}_a\hat{x}_i}\cr
& & - (\partial_0\phi_2-a_0\phi_1)\frac{1}{r}\epsilon_{aik}\hat{x}_k,
\\
F_{wi}^a &=& (\partial_2\phi_1+a_2\phi_2)\frac{\hat{\delta}_{ai}}{r} + (\partial_2a_1-\partial_1a_2) {\hat{x}_a\hat{x}_i} \cr & & - (\partial_2\phi_2-a_2\phi_1)\frac{1}{r}\epsilon_{aik}\hat{x}_k, \\
F_{0w}^a &=& (\partial_0a_2-\partial_2a_0)\hat{x}^a. 
\end{eqnarray}

With the Witten Ansatz, 
1+4 dimensional SU(2)$_f$ Yang-Mills sector of holographic QCD is reduced into 
a 1+2 dimensional Abelian Higgs theory on a curved space.
In fact, $S_{\rm 5YM}^{{\rm SU(2)}_f}$ is rewritten as 
\begin{eqnarray}
&&S_{\rm 5YM}^{{\rm SU(2)}_f} \cr
&=&-\kappa \int d^4x\ dw\ {\rm tr} \left[ 
\frac{1}{2} h(w) F_{\mu\nu} F^{\mu\nu} 
+ k(w)F_{\mu w} F^{\mu w} \right] \cr
&=& 4\pi\kappa \int^\infty_{-\infty} dt \int^\infty_0 dr \int^\infty_{-\infty} dw \biggl[ h(w) (|D_0\phi|^2 - |D_1\phi|^2) \cr
&-& k(w) |D_2\phi|^2 
- \frac{h(w)}{2r^2}(1-|\phi|^2)^2 \cr
&+& \frac{r^2}{2} \{ h(w) f_{01}^2 + k(w) f_{02}^2 - k(w) f_{12}^2 \} \biggr],
\label{eq:S5YM}
\end{eqnarray}
where the complex Higgs field $\phi(t,r,w) \in {\bf C}$, 
Abelian gauge field $a_\mu(t,r,w)$, its covariant derivative $D_\mu$, 
and field strength $f_{\mu\nu}$ in the Abelian Higgs theory are 
\begin{eqnarray}
\phi &\equiv& \phi_1+i\phi_2 \in {\bf C}, \
a_\mu \equiv (a_0,a_r,a_w), \cr 
D_\mu &\equiv& \partial_\mu - i a_\mu, \
f_{\mu\nu} \equiv \partial_\mu a_\nu-\partial_\nu a_\mu. \ \
\end{eqnarray}
Here, we have used $(0,1,2)=(t,r,w)$ for the index of 
1+2 dimensional coordinates. 
Note that the factor $k(w)$ appears in $D_2$ and $F_{12}$ 
associated with the index 2 ($w$), and otherwise the factor $h(w)$ appears. 

From this action, 
the static energy of the Yang-Mills part is obtained as
\begin{eqnarray}
&&E_{\rm 5YM}^{{\rm SU(2)}_f} [\phi(r,w), \vec a(r,w)]
\cr
&&= 4\pi\kappa \int^{\infty}_{0} dr \int^{\infty}_{-\infty} dw \biggl[ h(w) |D_1\phi|^2 + k(w) |D_2\phi|^2 \cr 
&& \qquad \qquad + \frac{h(w)}{2r^2}\{1-|\phi|^2\}^2 + \frac{r^2}{2} k(w) f_{12}^2 \biggr] \cr
&&= 4\pi \int^{\infty}_{0} dr r^2 \int^{\infty}_{-\infty} dw~{\cal E}^{{\rm SU(2)}_f}(r,w), 
\label{eqn:E_5YM}
\end{eqnarray}
with $\vec a=(a_1,a_2)=(a_r,a_w)$. 
This is similar with the 1+2 dimensional Ginzburg-Landau theory. 
As the different point, however, 
there appears a nontrivial metric of $r^2$ \cite{W77}
in addition to holographic background gravity of 
$h(w)$ and $k(w)$ in the $(r,w)$ half plane. 
To include all the gravitational effects exactly, 
we construct the lattice formalism, as shown in Appendix~A, 
and perform the numerical calculation for holographic QCD. 

Finally in the subsection, 
we investigate 
the topological density $\rho_B$
in the Witten Ansatz. 
For the SU(2)$_f$ gauge configuration in the Witten Ansatz, 
the topological density $\rho_B$ in $(x,y,z,w)$-space is found to be \cite{SH20}
\begin{eqnarray}
\rho_B &\equiv& \frac{1}{16\pi^2} {\rm tr}\bigl( F_{MN} \tilde{F}_{MN} \bigr) 
= \frac{1}{32\pi^2} \epsilon_{MNPQ} {\rm tr}\bigl( F_{MN} F_{PQ} \bigr)
\nonumber \\ 
&=&\frac{1}{8\pi^2r^2}
\{-i\epsilon_{ij}(D_i\phi)^*
D_j\phi+
\epsilon_{ij}
\partial_i a_j
(1-|\phi|^2)
\}
\nonumber \\ 
&=&\frac{1}{8\pi^2r^2}
\epsilon_{ij}
\partial_i
\{
\frac{1}{2i}(\phi^*\partial_j\phi-\phi\partial_j\phi^*)
+ a_j (1-|\phi|^2)
\}
\nonumber \\ 
&=&\frac{1}{8\pi^2r^2}\epsilon_{ij}\partial_i
\{a_j(1-|\phi|^2)+\partial_j \theta \cdot |\phi|^2\},
\label{eq:rho_B}
\end{eqnarray}
where $\theta\equiv {\rm arg}~\phi$ and 
Roman small letters 
$i, j$ take $(1,2)=(r,w)$.
The topological density $\rho_B$ is expressed as a total derivative. 
Since we consider $(x,y,z)$-rotationally symmetric system, 
$\rho_B$ is independent of the spatial direction $(\hat x, \hat y, \hat z)$. 
In fact,  $\rho_B$ takes an SO(3) rotationally symmetric form of $\rho_B(r,w)$ 
in the second line of Eq.~(\ref{eq:rho_B}).

\subsection{U(1) sector in holographic QCD}

In this subsection, we consider 
the U(1) sector in holographic QCD. 
Hereafter, the capital-letter index denotes 
the Euclidean spatial index as $M=x, y, z, w$. 
Also for the U(1) gauge field $\hat A$, 
we respect the spatial SO(3) rotational symmetry \cite{BS14,RSR14}
as in the Witten Ansatz, and impose 
\begin{eqnarray}
\hat A_i(t,x,y,z,w) &=& \hat a_r(t,r,w) \hat{x}_i, 
\end{eqnarray}
while $\hat A_0$ and $\hat A_w$ are treated to be arbitrary.
In this case, one finds $\hat F_{ij}=0$ and can take the $\hat a_r=0$ gauge, 
which simplifies $\hat A_i=0$.

Then, the $1/\lambda$-leading term $S_{\rm 5YM}^{\rm U(1)}$ 
for the U(1) gauge field is written as \cite{SH20}
\begin{eqnarray}
S^{\rm U(1)}
&=& 
\frac{\kappa}{2}
\int d^4x dw \{ h(w) \hat F_{0i}^2+k(w) \hat F_{0w}^2- k(w) \hat F_{iw}^2\} \cr
&=& \int d^4x dw \biggl[ 
\frac{1}{2}\hat A_0 K \hat A_0-\frac{\kappa}{2} k(w) (\partial_i \hat A_w)^2
 \cr 
 & & + ({\rm time\hbox{-}derivative~terms})
\biggr],
\end{eqnarray}
using the SO(3)-symmetric non-negative hermite kernel 
\begin{eqnarray}
K &\equiv& -\kappa \{ h(w) \partial_i^2 + \partial_w k(w) \partial_w \}
\cr
&=&-\kappa \{ h(w) \frac{1}{r^2}\partial_r r^2 \partial_r 
+ \partial_w k(w) \partial_w \}.
\end{eqnarray}
In this calculation, 
we take $\hat{A}_i=0$ 
using the gauge and rotational symmetry. 

We consider the CS term $S_{\rm CS}$ 
as the next leading order of the $1/\lambda$ expansion. 
For the static SO(3)-rotationally symmetric configuration in the $A_0=0$ gauge, 
the CS term $S_{\rm CS}$ in Eq.~(\ref{eq:CS}) 
is transformed as \cite{HSSY07,BS14,RSR14,SH20}
\begin{eqnarray}
S_{\rm CS} &=& \frac{N_c}{24\pi^2} \epsilon_{MNPQ} \int d^4x dw \left[ \frac{3}{8}\hat{A}_0{\rm tr}(F_{MN}F_{PQ}) \right. \nonumber \\
& & \left. - \frac{3}{2}\hat{A}_M{\rm tr}(\partial_0 A_NF_{PQ}) + \frac{3}{4}\hat{F}_{MN}{\rm tr}(A_0F_{PQ}) \right. \nonumber \\
& & \left. + \frac{1}{16}\hat{A}_0\hat{F}_{MN}\hat{F}_{PQ} - \frac{1}{4}\hat{A}_M\hat{F}_{0N}\hat{F}_{PQ} \right] \nonumber \\
&=& \frac{N_c}{2}\int d^4x dw~ \rho_B \hat A_0, \label{eqn: S CS}
\end{eqnarray}
up to total derivative. 
This is Coulomb-type interaction between 
the U(1) gauge potential $\hat A_0$ and 
the topological density 
$\rho_B \equiv \frac{1}{16\pi^2} {\rm tr}(F_{MN} \tilde{F}_{MN})$.

Then, the total U(1) action 
depending on the U(1) gauge field $\hat A$ is written as 
\begin{eqnarray}
&&S^{\rm U(1)} \equiv S^{\rm U(1)}+S_{\rm CS} \cr
&=& \int d^4x dw \biggl[ 
\frac{1}{2}\hat A_0 K \hat A_0 +\frac{N_c}{2}\rho_B \hat A_0 
-\frac{\kappa}{2} k(w) (\partial_i \hat A_w)^2
\biggr],~~~~~~
\label{eq:S^U(1)}
\end{eqnarray}
which leads to 
the field equations, 
\begin{eqnarray}
\vspace{-0.5cm}
K \hat A_0 +\frac{N_c}{2}\rho_B=0, \ \ \partial_i^2 \hat A_w=0.
\vspace{-0.5cm}
\label{eq:U(1)FE}
\end{eqnarray}
For the static configuration, 
the additional energy $E^{\rm U(1)}$ 
from U(1) sector $S^{\rm U(1)}$ 
is simply given by 
\begin{eqnarray}
&&E^{\rm U(1)}= -S^{\rm U(1)}/\int dt \cr
&=&\int d^3x dw \biggl[ 
-\frac{1}{2}\hat A_0 K \hat A_0 
-\frac{N_c}{2}\rho_B \hat A_0 
+\frac{\kappa}{2} k(w) (\partial_i \hat A_w)^2
\biggr]. \nonumber \\
\label{eq:E^U(1)org}
\end{eqnarray}

For the static case, $\hat A_w$ is dynamically isolated in the field equation (\ref{eq:U(1)FE}) 
and the last term is non-negative in the energy (\ref{eq:E^U(1)org}), 
and therefore we set $\hat A_w=0$, 
which satisfies the local energy-minimum condition. 
Then, one finds 
\begin{eqnarray}
E^{\rm U(1)}
=-\int d^3x dw \biggl[ 
\frac{1}{2}\hat A_0 K \hat A_0 
+\frac{N_c}{2}\rho_B \hat A_0 
\biggr].
\label{eq:E^U(1)}
\end{eqnarray}
For the SO(3) rotationally symmetric solution, 
we eventually obtain 
the static energy of the U(1) part: 
\begin{widetext}
\begin{eqnarray}
E^{\rm U(1)}[\rho_B(r,w), \hat A_0(r,w)]
&=&-4\pi \int_0^\infty dr r^2 \int_{-\infty}^{\infty}dw 
\biggl[ 
\frac{1}{2}\hat A_0(r,w) 
K \hat A_0(r,w) 
+\frac{N_c}{2}\rho_B(r,w) 
\hat A_0(r,w) 
\biggr] \cr
&=& -\int_0^\infty dr \int_{-\infty}^{\infty}dw  
\biggl[ 
\frac{1}{2}\hat A_0(r,w) 
\tilde{K} \hat A_0(r,w) 
+2\pi N_c\tilde{\rho}_B(r,w) 
\hat A_0(r,w) 
\biggr] \cr 
&=& 4\pi \int_0^\infty dr r^2 \int_{-\infty}^{\infty}dw~{\cal E}^{\rm U(1)}(r,w), 
\label{eq:E^U(1)rw}
\end{eqnarray}
\end{widetext}
using $\tilde \rho_B(r,w) \equiv r^{2}\rho_B(r,w)$ and 
the hermite kernel $\tilde K$ in $(r,w)$-space, 
\begin{eqnarray}
\tilde K \equiv 4\pi r^2 K
=-4\pi \kappa \{h(w) \partial_r r^2 \partial_r + r^2 \partial_wk(w) \partial_w\}.~~~~ 
\label{eq:kernel_rw}
\end{eqnarray}
For the numerical calculation of the U(1) sector, we mainly use this energy functional. 
(For another expression of $E^{\rm U(1)}$ 
after $\hat A^0$ path-integration, 
see Appendix~B.) 
 
Note again that, at the leading order of $1/\lambda$, 
the U(1) sector ($\hat A$) completely decouples 
with the SU(2)$_f$ sector ($\vec a$, $\phi$) 
because the leading term is only 
the Yang-Mills action (\ref{eq:5YM}).
However, at the next leading order of $1/\lambda$, 
the U(1) term affects the SU(2)$_f$ part through 
the CS term as Eq.~(\ref{eqn: S CS}). 

To summarize, the total energy $E$ comprises two parts,  
\begin{eqnarray}
    &&E[\phi(r,w), \vec a(r,w), \hat A_0(r,w)] \cr
    &=& E_{\rm 5YM}^{{\rm SU(2)}_f}[\phi, \vec a] 
    + E^{\rm U(1)}[\rho_B, \hat A_0], 
    \label{eqn:total energy}
\end{eqnarray} 
and the total energy density ${\cal E}(r,w)$ 
is written as 
\begin{eqnarray}
    {\cal E}(r,w)= {\cal E}^{{\rm SU(2)}_f}(r,w)
    + {\cal E}^{\rm U(1)}(r,w),
    \label{eqn:total energy density}
\end{eqnarray} 
and we have to deal with 
coupled field equations of the SU(2)$_f$ and U(1) sectors 
and will perform the numerical calculation 
in a consistent manner. 

\section{Vortex description of baryons}
\label{sec:vortex baryon}

In this section, 
we introduce vortex description of baryons  
in holographic QCD with 
applying the generalized Witten Ansatz. 
For a single baryon which is $(x,y,z)$-spatially rotational symmetric, 
applying the Witten Ansatz, we reduce the theory 
into a 1+2 dimensional Abelian Higgs theory in a curved space.
In the reduced theory, the holographic baryon is expressed as a two-dimensional topological object of an Abrikosov vortex.
We perform the numerical calculation 
of a $B=1$ solution of holographic QCD 
as the single ground-state baryon, 
using a fine and large lattice 
with spacing of 0.04 fm and size of 10 fm. 

\subsection{
Holographic baryons in Witten Ansatz
}

Large $N_c$ analyses of QCD indicate that 
explicit degree of freedoms are only mesons and glueballs, and 
baryons appear as solitons (topological objects) constructed with meson fields \cite{W79}. 
Also, holographic QCD based on large $N_c$ becomes an effective theory of mesons, 
and baryons appear as chiral solitons composed of meson fields 
in this framework \cite{SS05,NSK07}. 
Holographic QCD has four dimensional space $(x,y,z,w)$, and 
instantons naturally appear as relevant topological objects
in the four-dimensional space. 
The topological objects are physically identified 
as baryons in holographic QCD \cite{HSSY07}
and called holographic baryons. 

Remarkably, the Witten Ansatz generally converts 
the topological description from a four-dimensional instanton into a two-dimensional vortex \cite{W77}. 
Accordingly, the vortex number is interpreted as the baryon number in holographic QCD with the Witten Ansatz
\cite{SH20}. 

In fact, the baryon number $B$ or the Pontryagin index is written by 
a contour integral in the $(r,w)$-plane 
\begin{eqnarray}
B &=& \int d^3x dw~ \rho_B \cr
&=&\frac{1}{2\pi}
\int_0^\infty dr \int_{-\infty}^\infty dw
\epsilon_{ij}\partial_i
\{a_j(1-|\phi|^2)+\partial_j \theta \cdot |\phi|^2\}
\nonumber \cr
&=& \oint_{r \ge 0} d{\bf s} \cdot \{{\bf a}(1-|\phi|^2)+\nabla \theta \cdot |\phi|^2\} \\
&=& \oint_{r \ge 0} d{\bf s} \cdot \nabla \theta, 
\label{eq:bnum}
\end{eqnarray}
where $\oint_{r \ge 0}$ denotes the contour integral around 
the whole half-plane of $(r,w)$ with $r\ge 0$. 
To keep the energy (\ref{eqn:E_5YM})  finite, we have imposed 
the following boundary conditions: 
\begin{eqnarray}
&&|\phi(r=0, ^{\forall} \! w)| =1, \\
&&|\phi(^{\forall} \! r, w=\pm\infty)|= 
|\phi(r=\infty, ^{\forall} \! w)| = 1, 
\end{eqnarray}
at the edge of 
the $(r,w)$ half-plane.
Thus, the baryon number $B$ is converted into the vortex number in this formalism \cite{SH20}.

\subsection{Abrikosov vortex solution for a baryon in holographic QCD}

In this subsection, we numerically calculate 
a ground-state baryon of holographic QCD 
through the Abrikosov vortex description 
in the 1+2 dimensional U(1) Abelian Higgs theory \cite{SH20}.

With imposing the global condition of $B=1$, 
we numerically minimize the total energy $E[\phi, \vec a, \hat A_0]$ in  Eq.~(\ref{eqn:total energy}),  
which is equivalent to solving the equation of motion (EOM) 
of holographic QCD for the single ground-state baryon.

Regarding the two parameters $M_{\rm KK}$ and $\kappa$ 
in holographic QCD, 
we take $M_{\rm KK} \simeq$ 948~MeV, and $\kappa=7.46 \times10^{-3}$ 
to reproduce $f_\pi \simeq$ 92.4~MeV and $m_\rho \simeq$ 776~MeV \cite{SS05,NSK07}.
In this study, we have used 
the Kaluza-Klein unit of $M_{\rm KK}=1$.

For the numerical calculation on the $(r, w)$ plane, 
we adopt a fine and large-size lattice 
with spacing of $0.2~M_{\rm KK}^{-1} \simeq 0.04~{\rm fm}$ and 
the extension of $0 \le r \le 250$ and $-125 \le w \le 125$, 
that is, the system size is 
$250 \times 250$ grid corresponding to 
$(50~M_{\rm KK}^{-1})^2 \simeq (10~{\rm fm})^2$
in the physical unit. 
(In this numerical calculation, there appears subtle cancellation, 
and use of a coarse lattice might lead to an inaccurate result.
Also, the lattice size should be increased 
until the volume dependence of physical quantities disappears.)

On this lattice, 
starting from the 't~Hooft solution \cite{BPST, tH76} 
as a $B=1$ topological configuration, 
we numerically perform 
minimization of the total energy 
$E[\phi, \vec a, \hat A_0]$ 
keeping the topological charge 
by an iterative improvement, that is, 
many-time iterative local replacements  
of the field variable of 
$\phi$, $\vec a$ and $\hat A_0$. 
(See Appendix~A for the detail.)
During the update, 
the Higgs field $\phi(x)$ always has a zero point to ensure $B=1$, 
and the Higgs zero-point generally moves so as to realize the ground state.
In this way, we eventually 
obtain the holographic fields 
$\phi(r,w)$, $\vec a(r,w)$ and $\hat A_0(r,w)$ 
for the ground-state baryon
as the true solution of EOM in holographic QCD. 

For the confirmation of numerical calculations, 
we also consider another different method, 
as shown in Appendix~B. 
In this alternative method, 
we integrate out the U(1) gauge field $\hat A_0$ 
and update only $\phi(r,w)$ and $\vec a(r,w)$ 
on the lattice based on Eq.~(\ref{eqn:U(1) energy}). 
We have confirmed that both methods give 
the same numerical results for holographic baryons.

To visualize gauge and Higgs fields composing the Abrikosov vortex, 
we take the Landau gauge for the U(1) gauge degrees of freedom,  $\partial_i a_i(r,w) = 0$.
Of course, main results including the total energy 
are gauge invariant and never depend on any gauge choices. 

\begin{figure}[h]
    \begin{center}
    \includegraphics[width=90mm]{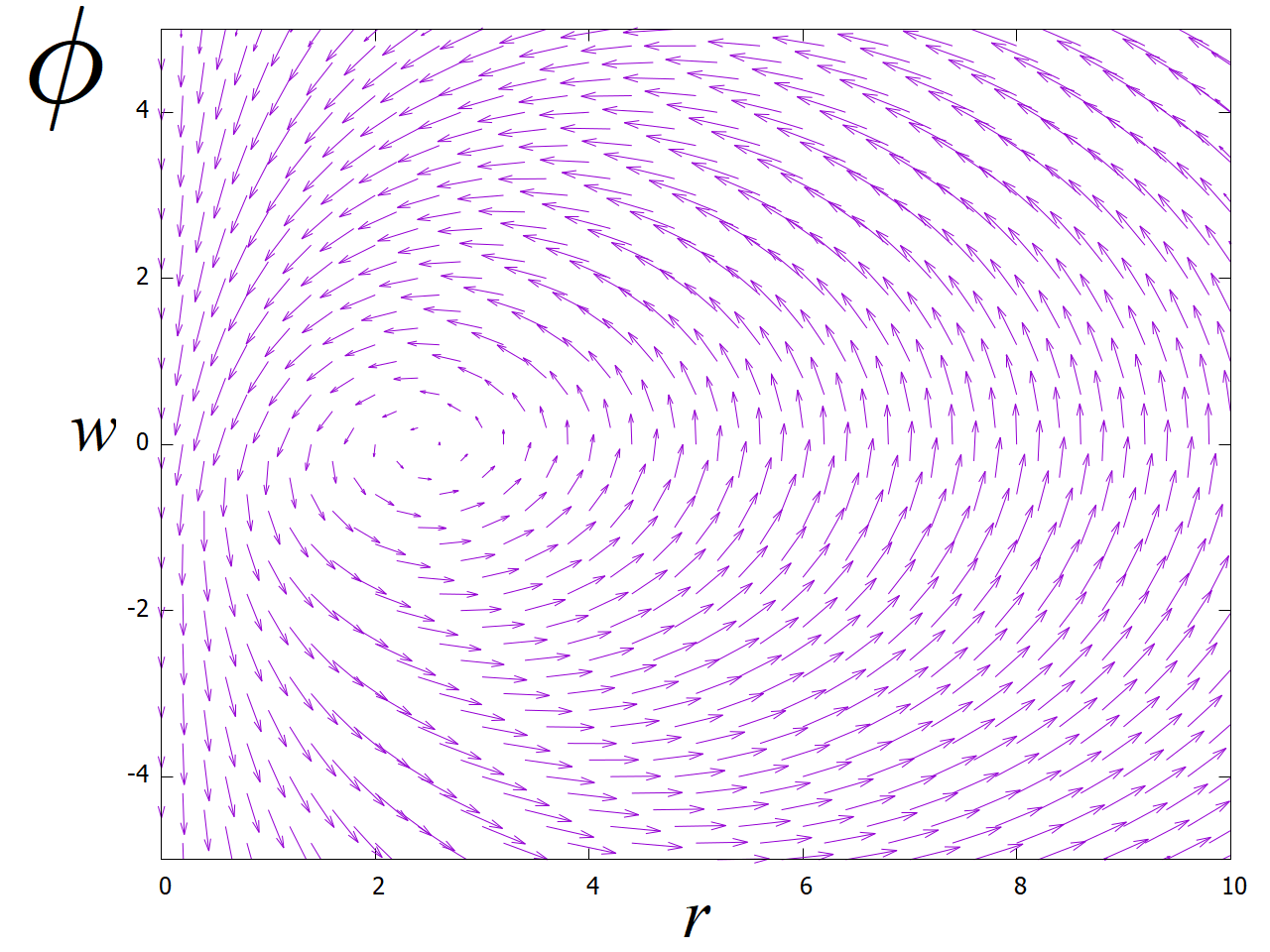}
    \includegraphics[width=90mm]{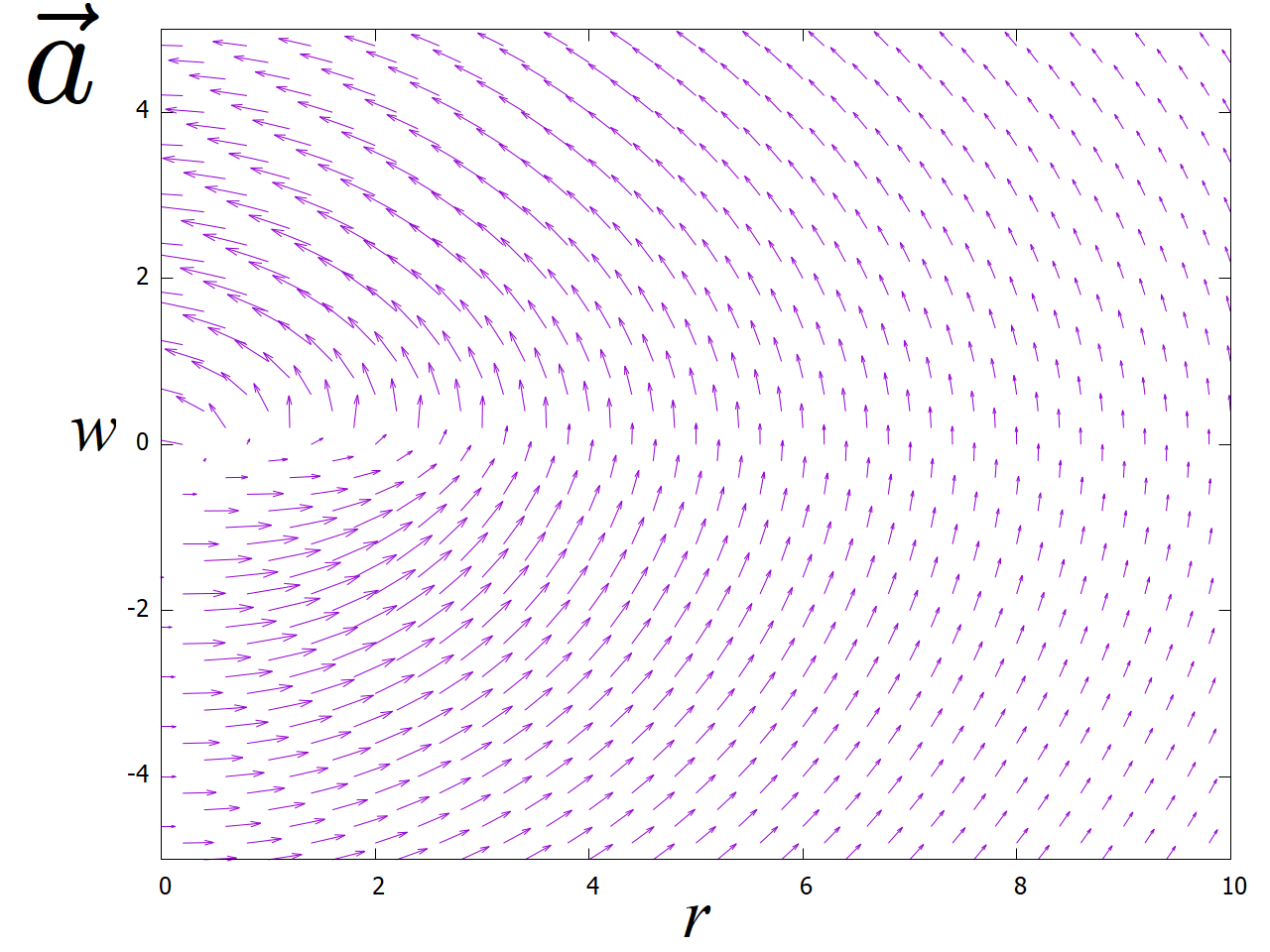}
    \end{center}
	\caption{Vortex description for 
	the single ground-state baryon: 
	the Higgs field $\phi(r,w) = ( {\rm Re}[\phi], {\rm Im}[\phi])$ (upper) and the Abelian gauge field 
	$a(r,w) = (a_1,a_2)$ (lower) in the Landau gauge. 
	The Kaluza-Klein unit of $M_{\rm KK}=1$ is used. 
	The Higgs field has a zero point 
    at $(r,w) \simeq (2.4, 0)$, 
	and its winding number around the zero point 
	is equal to the baryon number, that is, $B=1$.
 }
	\label{fig:phi a GS}
\end{figure}

\begin{figure}[h]
    \begin{center}
    \includegraphics[width=90mm]{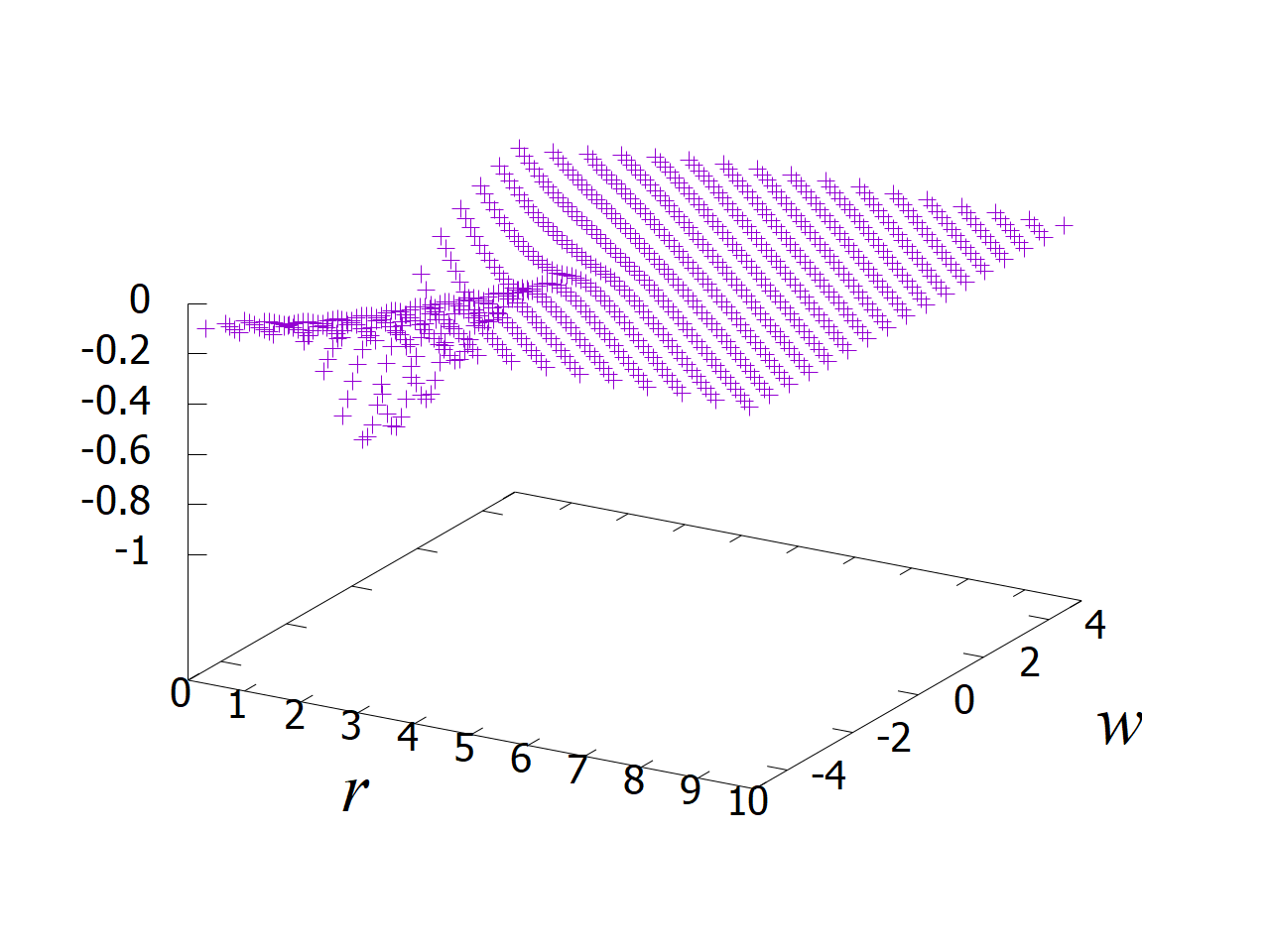}
    \end{center}
    \caption{
    Temporal component of the U(1) gauge field $\hat A_0(r,w)$ 
    in the ground-state holographic baryon 
    in the $M_{\rm KK}=1$ unit. 
    This U(1) gauge field $\hat A_0$ directly interacts with the topological density $\rho_B$ and 
    leads to a repulsive force between topological densities, 
    like the Coulomb interaction in QED.
    }
    \label{fig:aU1 GS}
\end{figure}

For the single ground-state baryon, 
we show field configurations $\phi(r,w)$ and $\vec a(r,w)$ in Fig.~\ref{fig:phi a GS}, 
and also the U(1) gauge field $\hat A_0(r,w)$ in Fig.~\ref{fig:aU1 GS}. 
The Higgs field $\phi(r,w)$ indicates a clear topological structure 
characterizing the Abrikosov vortex, 
which is mapped into an instanton \cite{W77} via the Witten Ansatz 
and physically means a baryon in holographic QCD \cite{BS14,SH20}. 
In accordance with the field equation (\ref{eq:U(1)FE}), 
$\hat A_0$ is localized 
around the non-vanishing topological density $\rho_B$, 
which will be shown in Fig.~\ref{fig:ene TpC GS rwd}. 

Quantitatively, unlike the initial 't~Hooft solution, 
Fig.~\ref{fig:phi a GS} no longer indicates 
the symmetry between $r$ and $w$ 
for the ground-state baryon solution in holographic QCD. 
For the ground-state baryon, 
both profiles of $\phi$ and $\vec a$ 
are found to be a little shrink in the $w$ direction,  
compared with four-dimensional spherical 't~Hooft solutions, which was also found 
in the previous numerical studies \cite{BS14,RSR14,SH20}.


\subsection{Properties of 
the ground-state baryon 
in holographic QCD}

\begin{figure}[h]
    \begin{center}
    \includegraphics[width=90mm]{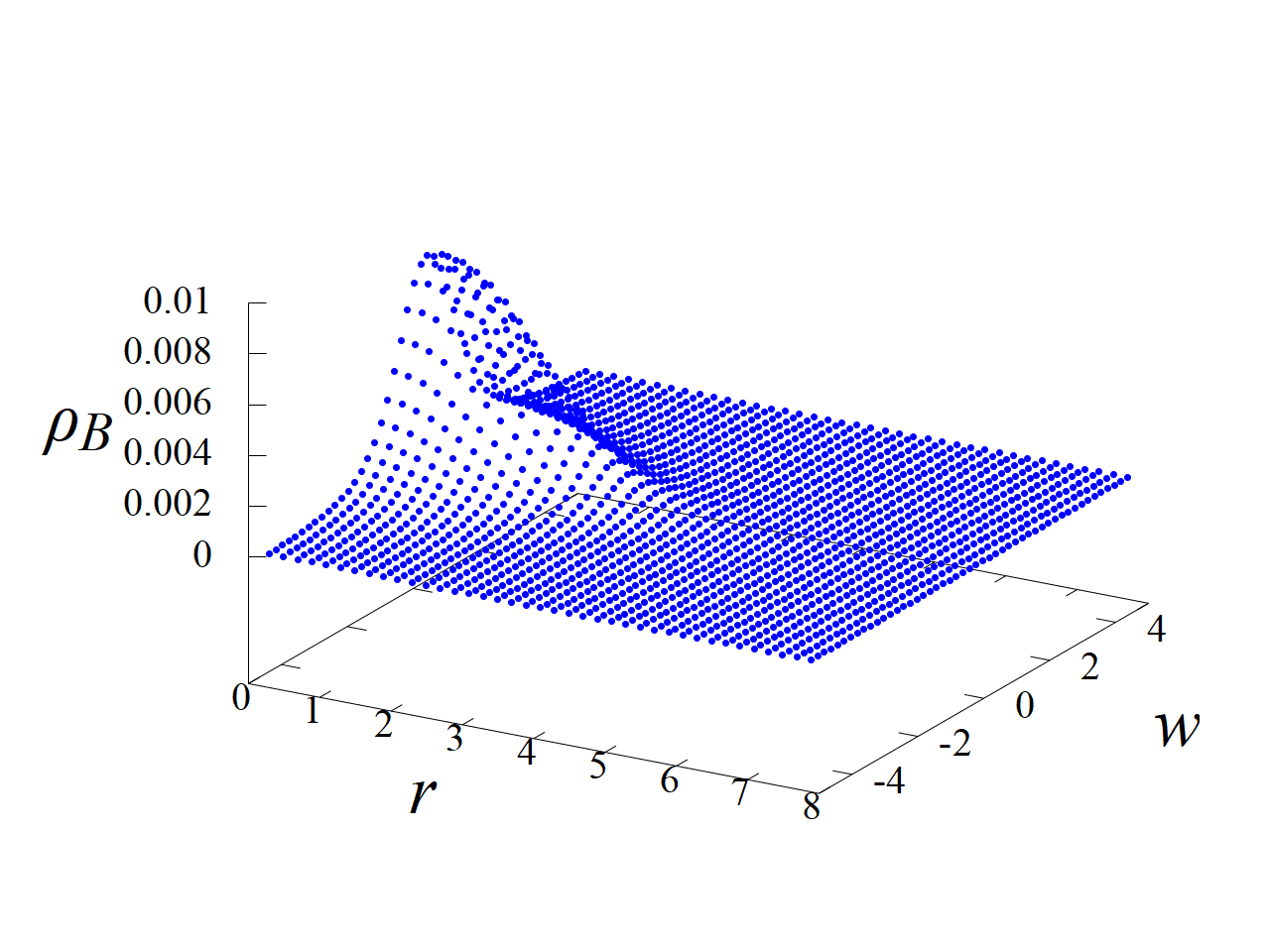}
    \includegraphics[width=90mm]{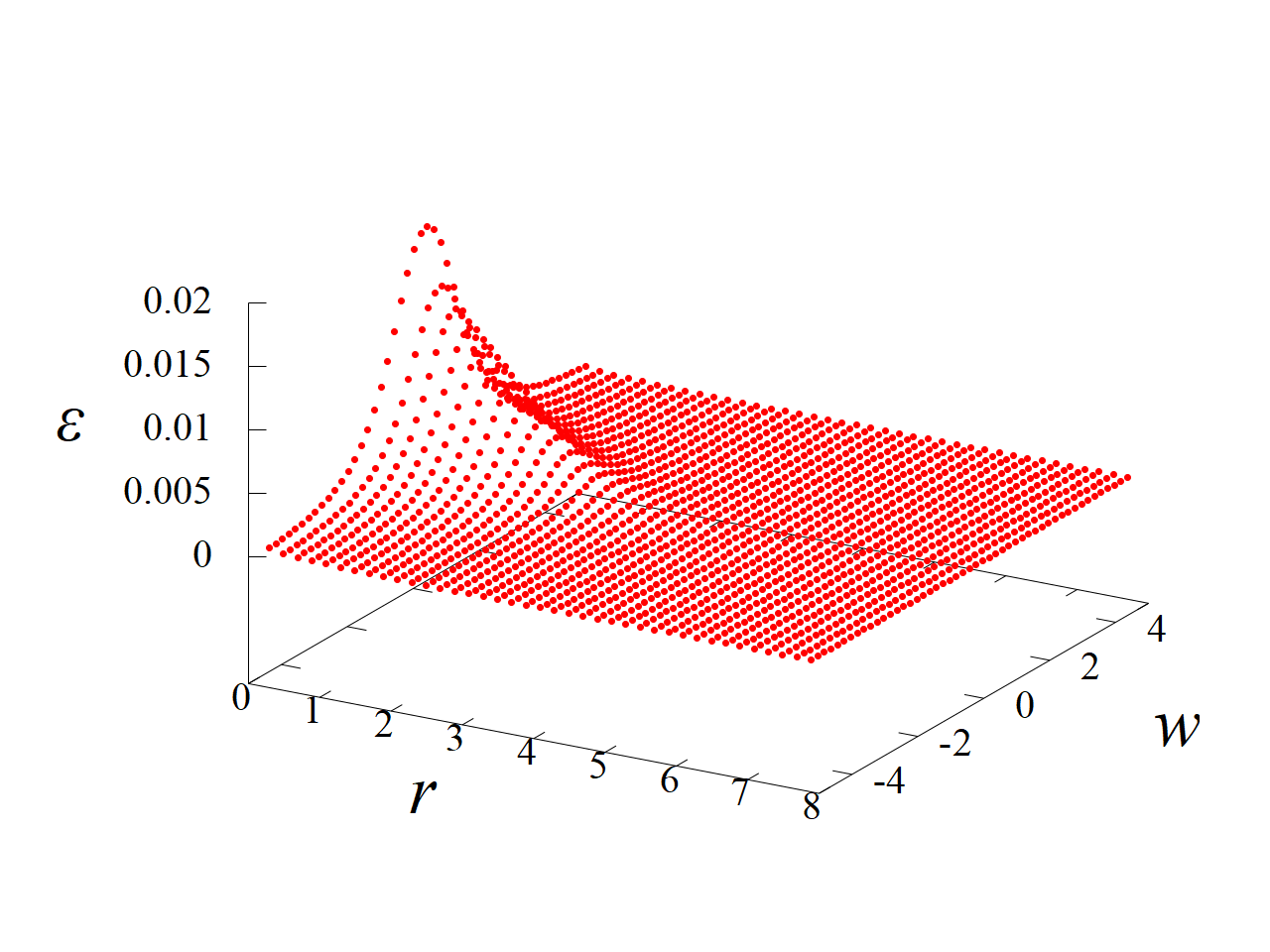}
    \end{center}
	\caption{Topological density $\rho_B(r,w)$ (upper) and total energy density ${\cal E}(r,w)$  (lower) for the ground-state solution of a single holographic baryon 
	in the $M_{\rm KK}=1$ unit.
	Both densities have a peak 
	around $(r, w)=(0, 0)$. 
	}
	\label{fig:ene TpC GS rwd}
\end{figure}

\begin{figure}[h]
    \begin{center}
    \includegraphics[width=90mm]{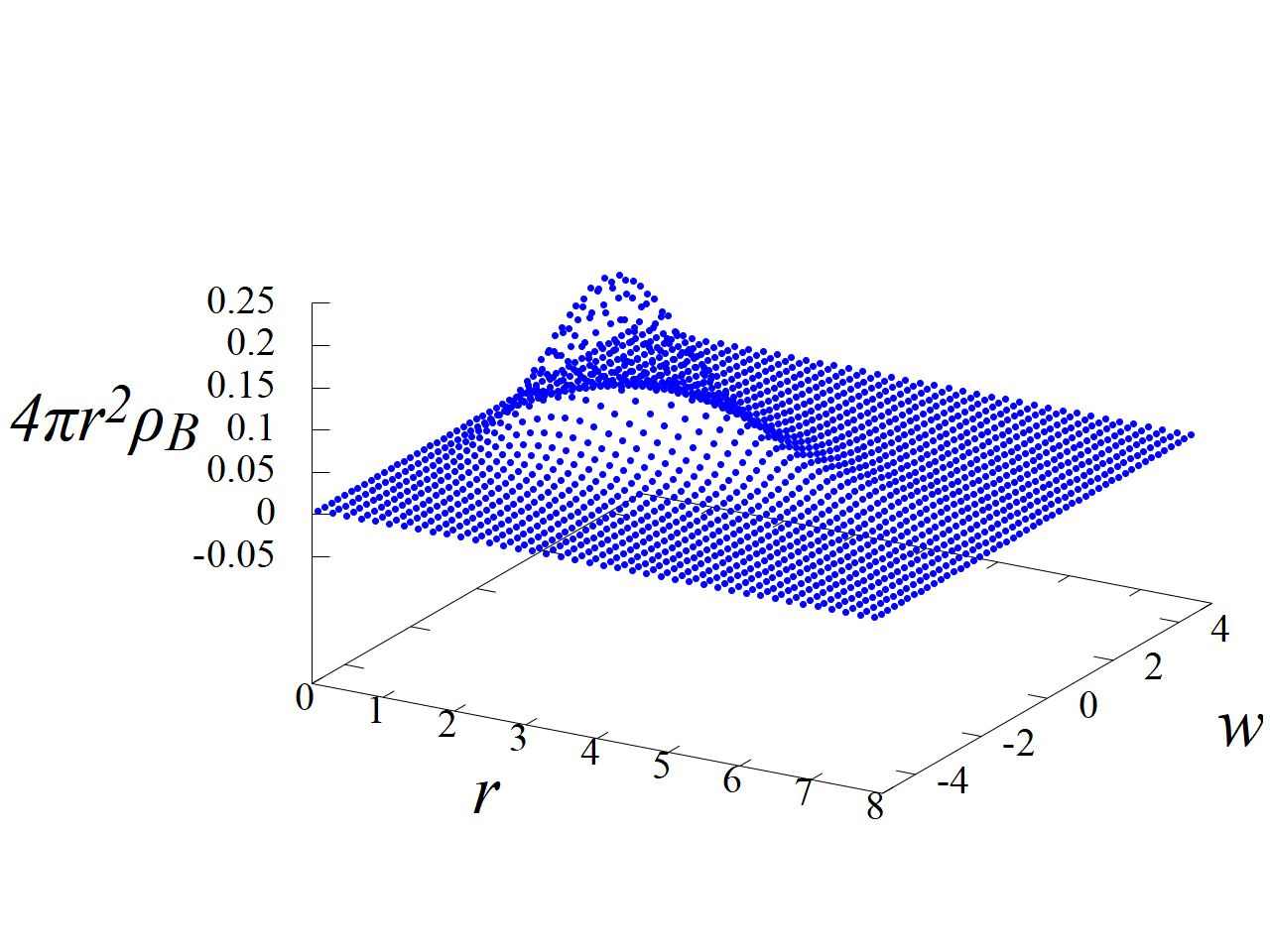}
    \includegraphics[width=90mm]{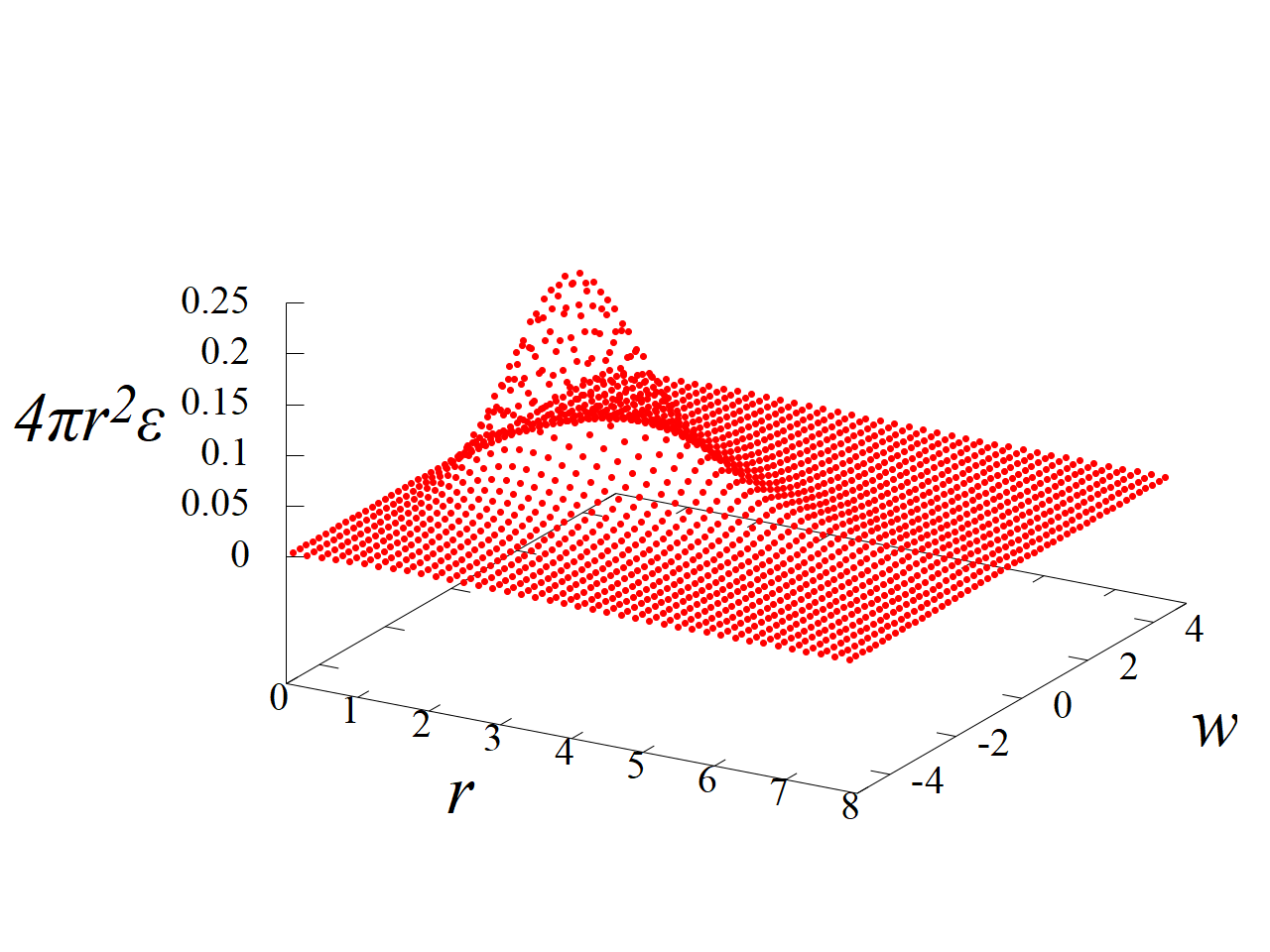}
    \end{center}
    \caption{
    $r^2$-multiplied topological and energy densities: 
    $4\pi r^2{\rho}_B(r,w)$ (upper) and 
    $4\pi r^2 {\cal E}(r,w)$ (lower) 
    for the ground-state solution of a single holographic baryon
    in the $M_{\rm KK}=1$ unit.
    Both figures have a peak 
    around $(r, w)\simeq (2, 0)$ near the vortex center, 
    which roughly controls the baryon size.
    }
    \label{fig:ene TpC GS rw}
\end{figure}

Now, we show the properties of 
the ground-state baryon 
in holographic QCD, which is numerically calculated by minimizing the total energy $E[\phi, \vec a, \hat A_0]$ in Eq.~(\ref{eqn:total energy}). 
In Fig.~\ref{fig:ene TpC GS rwd}, 
we show the topological density $\rho_B(r,w)$ 
and total energy density ${\cal E}(r,w)$ 
in the $(r,w)$-plane for the Abrikosov vortex solution in the 1+2 dimensional Abelian Higgs theory.
Both densities have a peak around $(r, w)=(0, 0)$ 
and are extended in both the $r$ and $w$ directions. 
We show in Fig.~\ref{fig:ene TpC GS rw} 
the densities multiplied by the integral measure factor $r^2$, 
i.e., $4\pi r^2 \rho_B(r,w)$ and $4\pi r^2 {\cal E}(r,w)$, 
for the ground-state baryon, 
since 
$\tilde \rho_B (r,w) \equiv r^2 \rho_B (r,w)$ 
is a primary variable in this numerical calculation, 
as shown in Eq.~(\ref{eq:kernel_rw}). 
The non-zero size of the baryon is 
due to the repulsive force from the CS term \cite{HSSY07}. 

In general, the integrated topological density is the baryon number $B$, 
and the baryon mass $M_B$ is given by integration of 
the total energy density ${\cal E}(r,w)$: 
\begin{eqnarray}
B&=&\int d^3x dw~ \rho_B(r,w) \cr
&=&4\pi \int_0^\infty dr r^2 \int_{-\infty}^\infty dw~ \rho_B(r,w),
\\
M_B&=&\int d^3x dw~ {\cal E}(r,w) \cr
&=&
4\pi \int_0^\infty dr r^2 \int_{-\infty}^\infty dw~ {\cal E}(r,w).
\label{eqn:baryon mass}
\end{eqnarray}
By integration over the extra coordinate $w$, 
we obtain the ordinary densities in a three-dimensional space, 
\begin{eqnarray}
\rho_B(r) &\equiv& \int_{-\infty}^\infty dw~\rho_B(r,w), \\
{\cal E}(r) &\equiv& \int_{-\infty}^\infty dw~{\cal E}(r,w). 
\label{eqn:ground energy}
\end{eqnarray}
The mass $M_B$ and size for the ground-state baryon are estimated as
\begin{eqnarray}
M_B &=& 
E_{\rm 5YM}^{{\rm SU(2)}_f}+E^{\rm U(1)}  
=\int d^3x ~{\cal E}(r) \cr
&\simeq& 1.25~{M_{\rm KK}} \simeq 1.19~{\rm GeV}, 
\label{M_B}
\end{eqnarray} 
\begin{eqnarray}
\sqrt{\langle r^2 \rangle}_{\rho_B} 
&\equiv& \sqrt{\frac{\int d^3x~ \rho_B(r) r^2}{\int d^3x~ \rho_B(r)}}
\cr 
&\simeq& 2.58 {M_{\rm KK}}^{-1} 
 \simeq 0.54\ {\rm fm},
 \label{r^2_rho_B}
\end{eqnarray} 
\begin{eqnarray}
\sqrt{\langle r^2 \rangle}_{\cal E} 
&\equiv& \sqrt{\frac{\int d^3x~ {\cal E}(r) r^2}{\int d^3x~ {\cal E}(r)}} 
\cr
&\simeq& 2.93\ {M_{\rm KK}}^{-1} 
\simeq 0.61\ {\rm fm}.
\label{r^2_rho_E}
\end{eqnarray} 

Here, some cautions are commented. 
When the self-dual BPS-saturated 't~Hooft solution~\cite{BPST,tH76} 
is simply used, as was done in Ref.~\cite{HSSY07}, 
the holographic baryon has an overestimated mass 
$M_B^{\rm BPS} \simeq 1.35~M_{\rm KK} \simeq 1.28~{\rm GeV}$ 
and a smaller radius $\sqrt{\langle r^2\rangle_{\rho_B}^{\rm BPS}} 
\simeq 2.2~M_{\rm KK}^{-1}\simeq 0.46~{\rm fm}$, 
as shown in Appendix~C. 
(Because of such an overestimation, 
a small value of $M_{\rm KK}=500~{\rm MeV}$ was adopted 
to adjust baryon masses in Ref.\cite{HSSY07}, 
which significantly differs from $M_{\rm KK} \simeq 1~{\rm GeV}$ 
for the meson sector.)
Another caution is the numerical accuracy, 
and fine and large lattices are to be used for the numerical calculation. 
Owing to a relatively coarse and small-size lattice, 
the numerical results in the previous paper \cite{SH20} 
include about 20\% error for the baryon mass and size. 

\begin{figure}[h]
    \begin{center}
    \includegraphics[width=90mm]{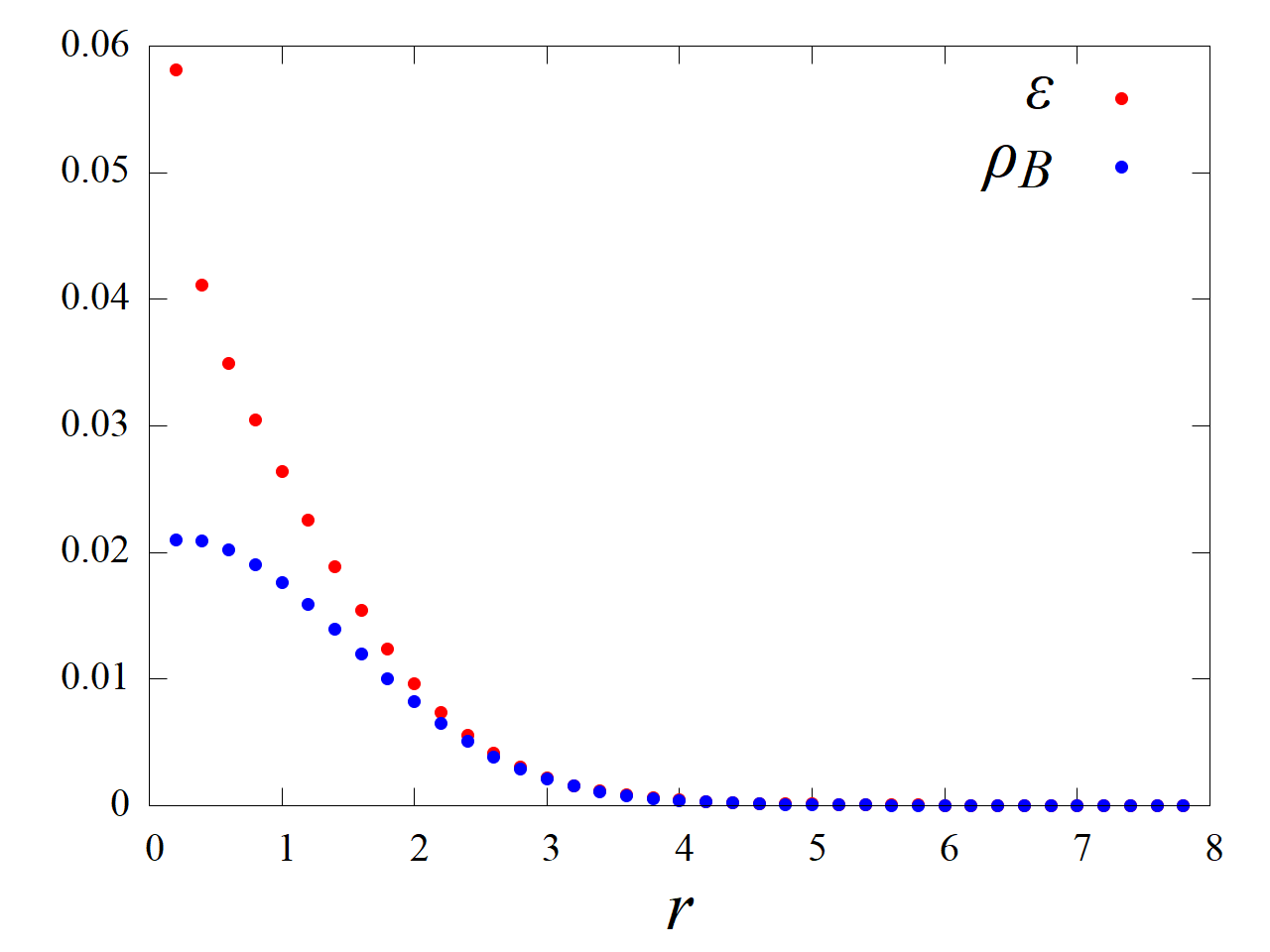}
    \includegraphics[width=90mm]{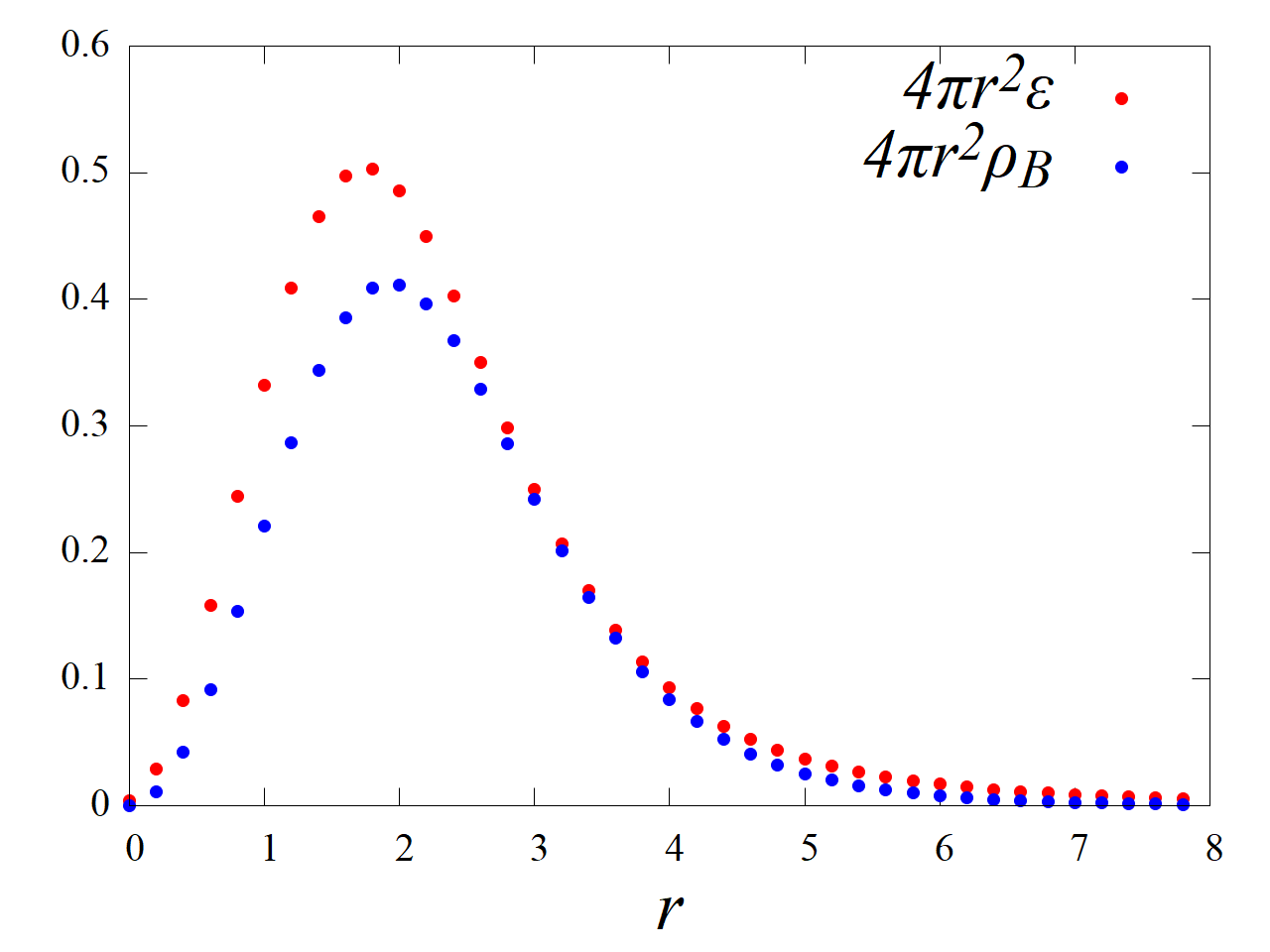}
    \end{center}
    \caption{Baryon density $\rho_B(r)$
	and total energy density ${\cal E}(r)$ for the ground-state baryon. 
	The lower panel shows $4\pi r^2 \rho_B(r)$ and $4\pi r^2 {\cal E}(r)$, including 
    the integral measure factor $r^2$. 
	Despite a significant difference 
    between $\rho_B(r)$ and ${\cal E}(r)$ 
    around the origin, 
    this difference is reduced by the $r^2$ multiplication. 
    Here, the $M_{\rm KK}=1$ unit is taken. 
    }
    \label{fig:ene TpC GS r}
\end{figure}

Now, we investigate spatial distribution of 
the baryon-number and energy densities 
for the ground-state baryon in holographic QCD. 
Figure~\ref{fig:ene TpC GS r} shows
the baryon-number density $\rho_B(r)$ 
(i.e. topological density) 
and total energy density ${\cal E}(r)$, 
and their $r^2$-multiplied values, 
$4\pi r^2 \rho_B(r)$ and $4\pi r^2 {\cal E}(r)$. 
One finds significant difference between the shapes of 
$\rho_B(r)$ and ${\cal E}(r)$ for the small $r$ region. 

Next, we investigate the energy contribution 
from the SU(2)$_f$ and U(1) parts, respectively. 
For the mass (total static energy)  
$M_B =E_{\rm 5YM}^{{\rm SU(2)}_f}+E^{\rm U(1)} 
$
of the ground-state holographic baryon, 
we obtain 
\begin{eqnarray}
E_{\rm 5YM}^{{\rm SU(2)}_f} &\simeq& 1.00 M_{\rm KK}
\simeq 0.95~{\rm GeV}, \\
E^{\rm U(1)} &\simeq& 0.25 M_{\rm KK}
\simeq 0.24~{\rm GeV}, 
\end{eqnarray}
and hence the SU(2)$_f$ contribution 
(leading order of $1/\lambda$ expansion) 
is found to be quantitatively dominant.

\begin{figure}[h]
    \begin{center}
    \includegraphics[width=90mm]{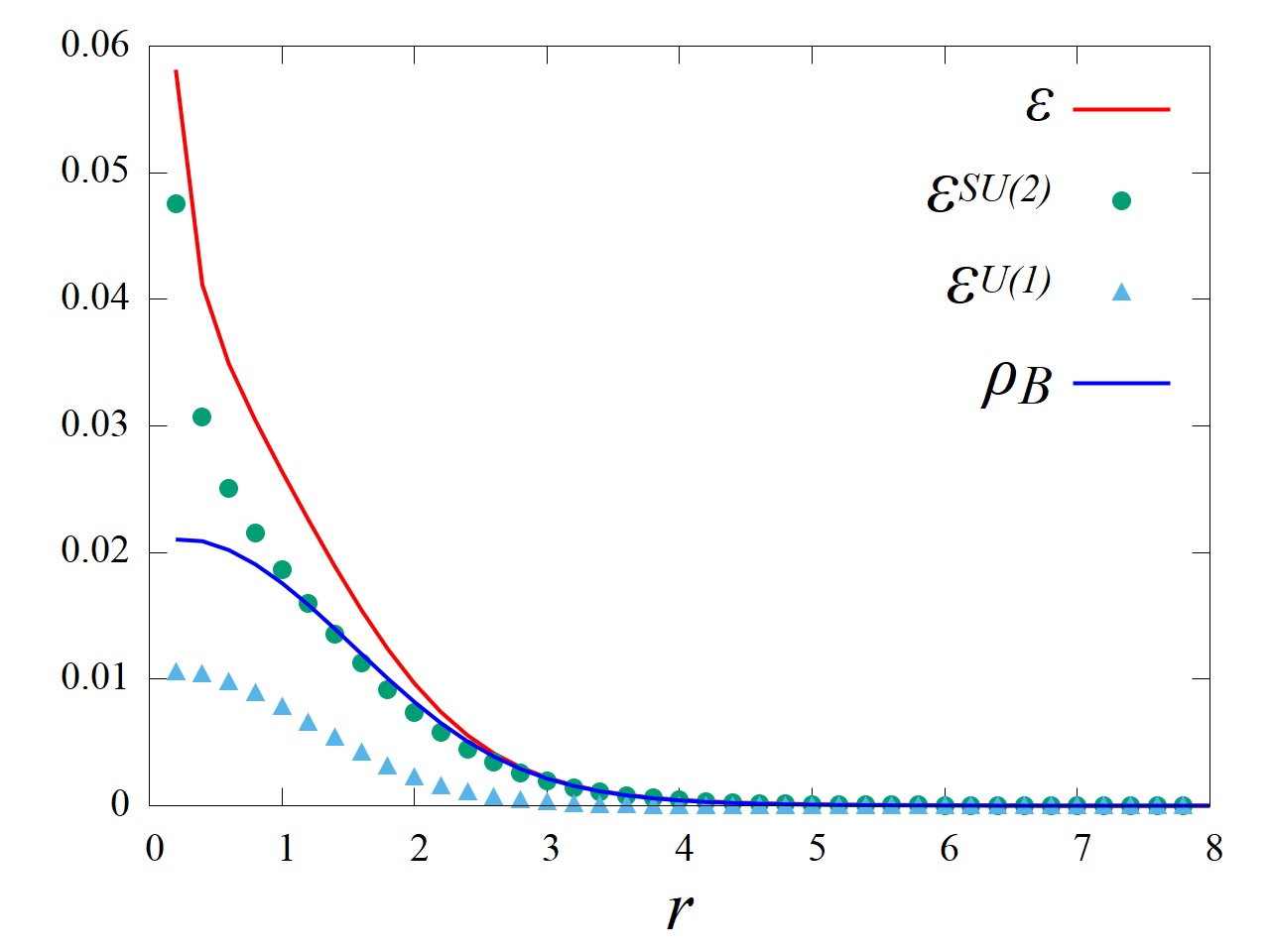}
    \includegraphics[width=90mm]{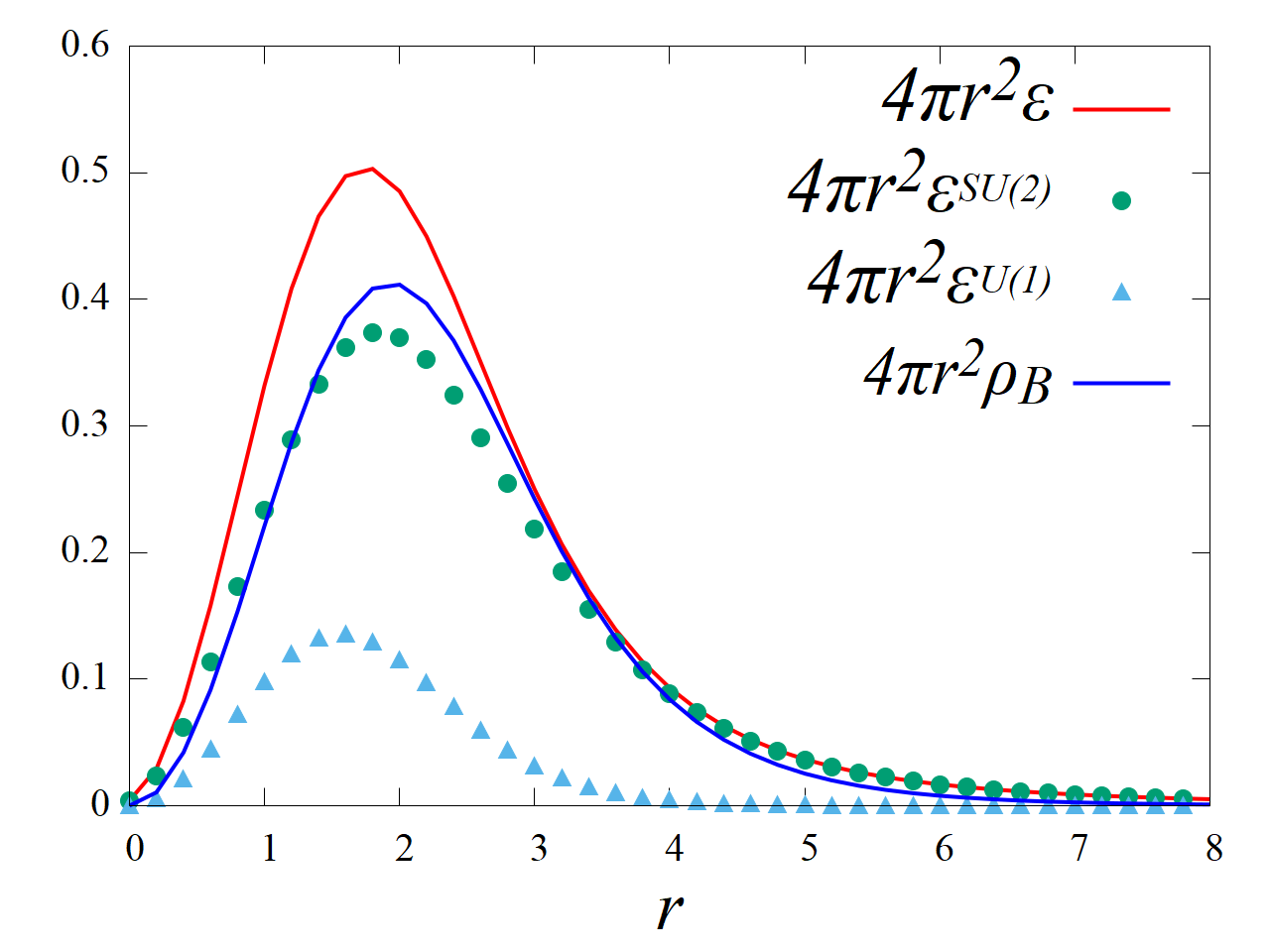}
    \end{center}
    \caption{
    SU(2)$_f$ energy density ${\cal E}^{{\rm SU(2)}_f}(r)$ and U(1) energy density ${\cal E}^{\rm U(1)}(r)$ of the ground-state baryon, together with the total energy density 
    ${\cal E}(r)$ and baryon density $\rho_B(r)$, 
    denoted by the solid lines. 
    The lower panel shows $r^2$-multiplied values, $4\pi r^2{\cal E}(r)$, $4\pi r^2{\cal E}^{{\rm SU(2)}_f}(r)$, $4\pi r^2{\cal E}^{\rm U(1)}(r)$ and $4\pi r^2\rho_B(r)$. 
    Here, the $M_{\rm KK}=1$ unit is taken. 
    }
    \label{fig:rho energy 4 types}
\end{figure}

As spatial distribution, 
the SU(2)$_f$ and U(1) energy densities are expressed as 
\begin{eqnarray}
&&r^2{\cal E}^{{\rm SU(2)}_f} (r) = \kappa \int^{\infty}_{-\infty} dw \biggl[ h(w) |D_1\phi|^2 + k(w) |D_2\phi|^2 \cr 
&& \qquad \qquad + \frac{h(w)}{2r^2}\{1-|\phi|^2\}^2 + \frac{r^2}{2} k(w) f_{12}^2 \biggr], \\
\label{eqn:E_5YM (r)}
&&{\cal E}^{\rm U(1)}(r) 
= -\int_{-\infty}^{\infty} dw \cr 
&& \biggl[ 
\frac{1}{2}\hat A_0(r,w) 
 K \hat A_0(r,w) 
+\frac{N_c}{2} \rho_B(r,w) 
\hat A_0(r,w) 
\biggr].
\label{eq:E^U(1) (r)}
\end{eqnarray}
Figure~\ref{fig:rho energy 4 types} 
shows 
SU(2)$_f$ energy density ${\cal E}^{{\rm SU(2)}_f}(r)$ and U(1) energy density ${\cal E}^{\rm U(1)}(r)$ of the ground-state baryon, together with the total energy density ${\cal E}(r)$ and baryon density $\rho_B(r)$. 

The dominant contribution is the SU(2)$_f$ part, 
and the total value ${\cal E}(r)$ is approximated by 
the SU(2)$_f$ energy density ${\cal E}^{{\rm SU(2)}_f}(r)$, 
which seems to be consistent with $1/\lambda$ expansion.
In particular, ${\cal E}^{{\rm SU(2)}_f}(r)$ and ${\cal E}(r)$ has the same slope 
at the origin $r=0$ and show enhancement for small $r$ region, 
although it is masked by the integral measure $r^2$. 
The shape of ${\cal E}^{\rm U(1)}(r)$ seems to follow $\rho_B(r)$,  
reflecting the direct coupling between $\hat{A}_0$ and $\rho_B$ 
in the CS term (\ref{eqn: S CS}).

Finally, we investigate self-duality breaking of the holographic baryon. 
The different shape between the topological and energy densities 
originates from the background gravity, $h(w)$ and $k(w)$, 
and presence of the CS term. 
In the case of $h(w)=k(w)=1$ without the CS term, 
the instanton solution has exact self-duality of 
$F_{MN} = \tilde F_{MN}$, 
leading to $\rho_B(r,w) \propto {\cal E}(r,w)$. 
Hence, the functional forms of $\rho_B(r)$ and ${\cal E}(r)$ 
are forced to be the same. 
In fact, between the shapes of $\rho_B(r)$ and ${\cal E}(r)$, 
the similarity indicates the self-dual tendency, 
and the different point indicates self-duality breaking. 

For the single baryon case $B=1$, 
we introduce the self-duality breaking parameter defined by 
\begin{eqnarray}
    \Delta_{\rm DB} &\equiv& 
    \frac{
    \int d^3x dw~{\rm tr}( F_{MN}F_{MN} - F_{MN}\tilde{F}_{MN}) }{ \int d^3x dw~ {\rm tr} (F_{MN}\tilde{F}_{MN})} \cr
    &=&
    \frac{1}{32\pi^2} \int d^3x dw~{\rm tr}
    ( F_{MN} - \tilde{F}_{MN})^2, 
    \label{eq:self-duality breaking parameter}
\end{eqnarray}
which is normalized by the topological quantity. 
This is non-negative and becomes zero only 
in the exact self-dual case.
The self-duality breaking of the ground-state baryon is 
found to be ${\Delta}_{\rm DB} \simeq 0.17$, 
which is non-zero but seems small. 
The small value might indicate that 
the true holographic configuration is close to be self-dual. 
Then, the ground-state baryon might be approximated 
by the self-dual 't~Hooft instanton in holographic QCD.

\section{Size-dependence of a holographic baryon}\label{sec:size-dependence}

As above mentioned, by using 
the Witten Ansatz, 1+4 dimensional 
holographic QCD is reduced into 
a 1+2 dimensional Abelian Higgs theory. 
Accordingly, the topological description 
of a baryon is changed from an instanton 
to a vortex. 
In the previous section, 
we have numerically obtained the ground-state solution in holographic QCD, 
where the baryon size is automatically determined by minimizing the total energy. 

Now, let us consider a holographic baryon with various size. 
Note that the size is originally one of the moduli of an instanton in a flat space, 
and different size baryons are to be 
degenerate in holographic QCD, 
if one sets $h(w)=k(w)=1$ and 
neglects the CS term. 
In the real holographic QCD,
the size parameter is no more modulus, 
according to the background gravity and CS term. 
Nevertheless, the size might behave as a quasi-modulus in the holographic baryon, resulting in physical appearance of a soft vibrational mode as a low-lying excitation. 
Then, we investigate 
a baryon with various size 
and size dependence of the energy 
in holographic QCD in this section. 

First, we consider the ordinary Yang-Mills theory 
and examine the moduli relation between an instanton and a vortex 
in the Witten Ansatz, 
as was originally shown by Witten \cite{W77}. 

Next, we proceed to the holographic baryon, and consider how 
to obtain an arbitrary-size baryon as a solution of holographic QCD, 
and investigate the size dependence of 
the baryon mass. 

\subsection{Instanton-vortex correspondence in Yang-Mills theory}

In the four-dimensional Euclidean Yang-Mills theory, 
there exist topological solutions, instantons, 
where Euclidean time is necessary. 
A single instanton solution (BPST-'t~Hooft solution~\cite{BPST,tH76}) 
is written as 
\begin{eqnarray}
	A_\mu(x) = - \eta_{\mu\nu}^a\tau^a \frac{x^\nu}{(x-X)^2+R^2}, \label{eq:BPST 4dim}
\end{eqnarray}
where the instanton center locates at $x^\mu=X^\mu$, 
and $R$ denotes the instanton size. 
Together with color rotation, the location $X^\mu$ and the size $R$ are known as moduli, 
and they represent degrees of freedom for this topological solution, that is, their values do not affect the Yang-Mills action. 
Here, $\eta_{\mu\nu}^a$ denotes the 't~Hooft symbol \cite{tH76} defined by
\begin{eqnarray}
\eta_{\mu\nu}^a
=-\eta_{\nu\mu}^a=\Bigg\{\begin{array}{l}\epsilon_{a\mu\nu}~~~~{\rm for}~~\mu,\nu=1,2,3\\-\delta_{a\nu}~~~{\rm for}~~\mu=4 \\ ~~\delta_{a\mu}~~~{\rm for}~~\nu=4. 
\end{array} 
\end{eqnarray}
Taking its center $X^\mu=( \vec{0}, T)$, there is SO(3) rotational symmetry in $(x,y,z)$-space, and the Witten Ansatz is applicable. 
With the Witten Ansatz, 
the four-dimensional Yang-Mills theory is reduced into 
a two-dimensional Abelian Higgs theory, and 
this instanton can be described by a single vortex. 

To understand the relation 
between an instanton and a vortex, 
let us consider the 't~Hooft solution in the form of the Witten Ansatz. 
This represents a single instanton solution 
and is rewritten as 
\begin{eqnarray}
	A_i^a &=& \frac{2r}{r^2+(t-T)^2+R^2}\epsilon_{iaj} \hat{x}_j \cr 
	&& - \frac{2(t-T)}{r^2+(t-T)^2+R^2}(\hat{\delta}_{ia}+\hat{x}_i\hat{x}_a), \\
	A_t^a &=& \frac{2r}{r^2+(t-T)^2+R^2} \hat{x}_a . 
\end{eqnarray}
By comparing the functional form 
with Eqs.~(\ref{eqn:Original Witten Ansatz 1}) and (\ref{eqn:Original Witten Ansatz 2}), 
SU(2) gauge fields can be converted into 
the fields of the reduced Abelian Higgs theory, 
and their forms are obtained as 
\begin{eqnarray}
&&    (\phi_1,\phi_2+1) =\frac{2r}{r^2+(t-T)^2+R^2}  
    \left( {-(t-T)}, {r} \right), 
    \\ 
&& a_1  = \frac{-2(t-T)}{r^2+(t-T)^2+R^2},\ a_2 = \frac{2r}{r^2+(t-T)^2+R^2}.~~~~~~
\end{eqnarray}
The complex Higgs field 
$\phi=({\rm Re} \phi, 
\ {\rm Im} \phi) = (\phi_1, \phi_2)$ 
takes zero at $(r,t) = (R,T) \equiv \zeta$. 
The vortex number is counted as the zero-point number in the Higgs field $\phi$ 
in the $(r,t)$-plane. 
Now, there is one zero point at $\zeta$, and this configuration represents a single vortex. 

Thus, in the Witten Ansatz, 
the vortex corresponding to a single instanton 
has a zero point $\zeta$ of the Higgs field $\phi$. 
Remarkably, 
this Higgs-field zero point $\zeta$ relates to 
instanton parameters \cite{W77}, 
\begin{eqnarray}
    \zeta =(\zeta_{r}, \zeta_{t})= (R,\ T). \label{eqn:moduli relation}
\end{eqnarray}
In fact, the instanton size $R$ and Euclidean fourth-coordinate $T$ 
of the instanton center correspond to this zero point $\zeta$ in the Higgs field $\phi$. 
In this configuration, the topological density is written by 
\begin{eqnarray}
    \rho_B = \frac{6}{\pi^2} \frac{R^4}{[r^2+(t-T)^2+R^2]^4}.
\end{eqnarray}
The topological density localizes 
around $(r,t)=(0,T)$, 
and its extension is about the size parameter $R$, 
which reflects the instanton size. 

\subsection{Various-size baryon mass in holographic QCD}

In this subsection, 
using the above correspondence (\ref{eqn:moduli relation})
between the Higgs zero-point $\zeta$ and the instanton size $R$ in the Witten Ansatz, 
we try to control the baryon size by changing 
the location of the zero point $\zeta$ in the Higgs field $\phi$ in holographic QCD. 
Using this new viewpoint, 
we obtain various sizes of holographic baryons 
and investigate size dependence 
of the single baryon energy. 

To begin with, we reformulate the above argument 
for 1+4 dimensional holographic QCD.
We recall different points 
from the ordinary Yang-Mills theory. 
First, holographic QCD already has four-dimensional Euclidean 
space of $(x,y,z,w)$, and the instanton can be naturally defined 
on this space including the fourth spatial coordinate $w$,  
without use of Euclidean process.
Second, there appear the gravitational factor, $h(w)$ and $k(w)$, 
and the CS term, which results in a repulsive interaction 
among the baryon density $\rho_B$ in holographic QCD. 
Owing to these effects, the 't~Hooft solution is no longer the exact solution but an approximate one in holographic QCD. 

Therefore, we use the 't~Hooft solution as 
a starting point for the $B=1$ configuration, 
and search for the true solution of holographic QCD 
by a numerical iterative method 
with keeping the topological property unchanged, 
similarly in Sec.~IV. 

In 1+4 dimensional holographic QCD, taking the $A_0=0$ gauge, 
we use a holographic version of 
the 't~Hooft solution (\ref{eq:BPST 4dim}),  
as a starting point of the topological configuration. 
Here, we locate the instanton center at the four-dimensional 
spatial origin $(x,y,z,w)=(0,0,0,0)$ for symmetry of 
$(x,y,z)$-spatial rotation and $w$-reflection,
and then the initial configuration is set to be 
\begin{eqnarray}
	A_0 &=& 0, \\
	A_M &=& - \eta_{MN}\frac{x^N}{x^2+R^2} 
 \label{eqn:'t hooft soltion HQCD},
\end{eqnarray}
which can be rewritten as
\begin{eqnarray}
	A_i^a &=& \frac{2r}{r^2+w^2+R^2}\epsilon_{iaj} \hat{x}_j \cr 
	&& - \frac{2w}{r^2+w^2+R^2}(\hat{\delta}_{ia}+\hat{x}_i\hat{x}_a), \\
	A_w^a &=& \frac{2r}{r^2+w^2+R^2} \hat{x}_a. 
\end{eqnarray}
For the $(x,y,z)$-rotational symmetric system, 
through the Witten Ansatz, 
the non-Abelian gauge fields are 
converted into 
the Higgs field $\phi$ 
and Abelian gauge field $\vec a$, 
and they are expressed as
\begin{eqnarray}
&&    (\phi_1,\phi_2+1) =\frac{2r}{r^2+w^2+R^2} 
    \left( -w, r \right), \label{eqn:phi HQCD}\\ 
&& a_1  = \frac{-2w}{r^2+w^2+R^2},\ a_2 = \frac{2r}{r^2+w^2+R^2}. 
\end{eqnarray}
In Eq.~(\ref{eqn:phi HQCD}), 
the zero point $\zeta$ of the Higgs field $\phi$ 
is found to locate at 
$    \zeta =(\zeta_{r}, \zeta_{w})= (R,\ 0),$ 
and this $\zeta_r$ determines the initial size of the holographic baryon.

From this initial configuration, similarly in Sec.~IV, 
we numerically search for the single baryon solution in holographic QCD 
by minimizing the total energy $E[\phi, \vec a, \hat A_0]$, 
with fixing the Higgs-zero location 
\begin{eqnarray}
\zeta=(\zeta_r,\zeta_w)=(R,0),
\label{eqn:moduli relation HQCD}
\end{eqnarray}
where the former $R$ corresponds to 
the instanton/baryon size. 
In fact, to consider various size baryons, 
we here use the correspondence between the instanton/baryon size 
and zero point $\zeta$ in the Higgs field $\phi$ composing 
the Abrikosov vortex \cite{W77}.
Note that the Higgs field must be zero 
at the center of the Abrikosov vortex 
to realize the finite energy, 
and therefore the Higgs field $\phi$ 
inevitably has a zero-point $\zeta$ 
in the presence of the vortex 
corresponding to $B=1$ in holographic QCD. 
Here, to change the zero-point location $\zeta_r=R$ 
corresponds to a baryon-size change, 
and its various changes lead to different-size baryons. 

In summary, to obtain various-size baryon solutions, 
we fix this zero-point location $\zeta=(\zeta_r,\zeta_w)=(R,0)$ of 
the Higgs field $\phi$ as a boundary condition 
and minimize the total energy $E$ numerically. 
When the total energy $E[\phi, \vec a, \hat A_0]$ 
is minimized against arbitrary local variation 
of $\phi$, $\vec a$ and $\hat A_0$,   
these fields satisfies the EOM of holographic QCD.

For the numerical calculation on the $(r, w)$ plane, 
we use the same lattice used in Sec.~IV, that is, 
a fine and large-size lattice 
with spacing of $0.2 M_{\rm KK}^{-1} \simeq 0.04~{\rm fm}$ and 
the extension of $0 \le r \le 250$ and $-125 \le w \le 125$, 
of which physical size is  
$(50~M_{\rm KK}^{-1})^2 \simeq (10~{\rm fm})^2$. 

After an iterative improvement of holographic fields 
with the constraint of the Higgs zero-point location, 
we eventually obtain the Higgs and Abelian gauge fields, 
\begin{eqnarray}
\phi(r, w; R), \quad \vec a(r, w; R), \quad \hat{A}_0(r, w; R)
\end{eqnarray}
for the holographic baryon corresponding to the instanton size $R$.
Note again that the presence of a zero point of the Higgs field 
indicates a $B=1$ configuration, and 
the holographic fields obeying the local energy minimum condition 
also satisfy the EOM of holographic QCD. 

\begin{figure}[h]
    \begin{center}
    \includegraphics[width=90mm]{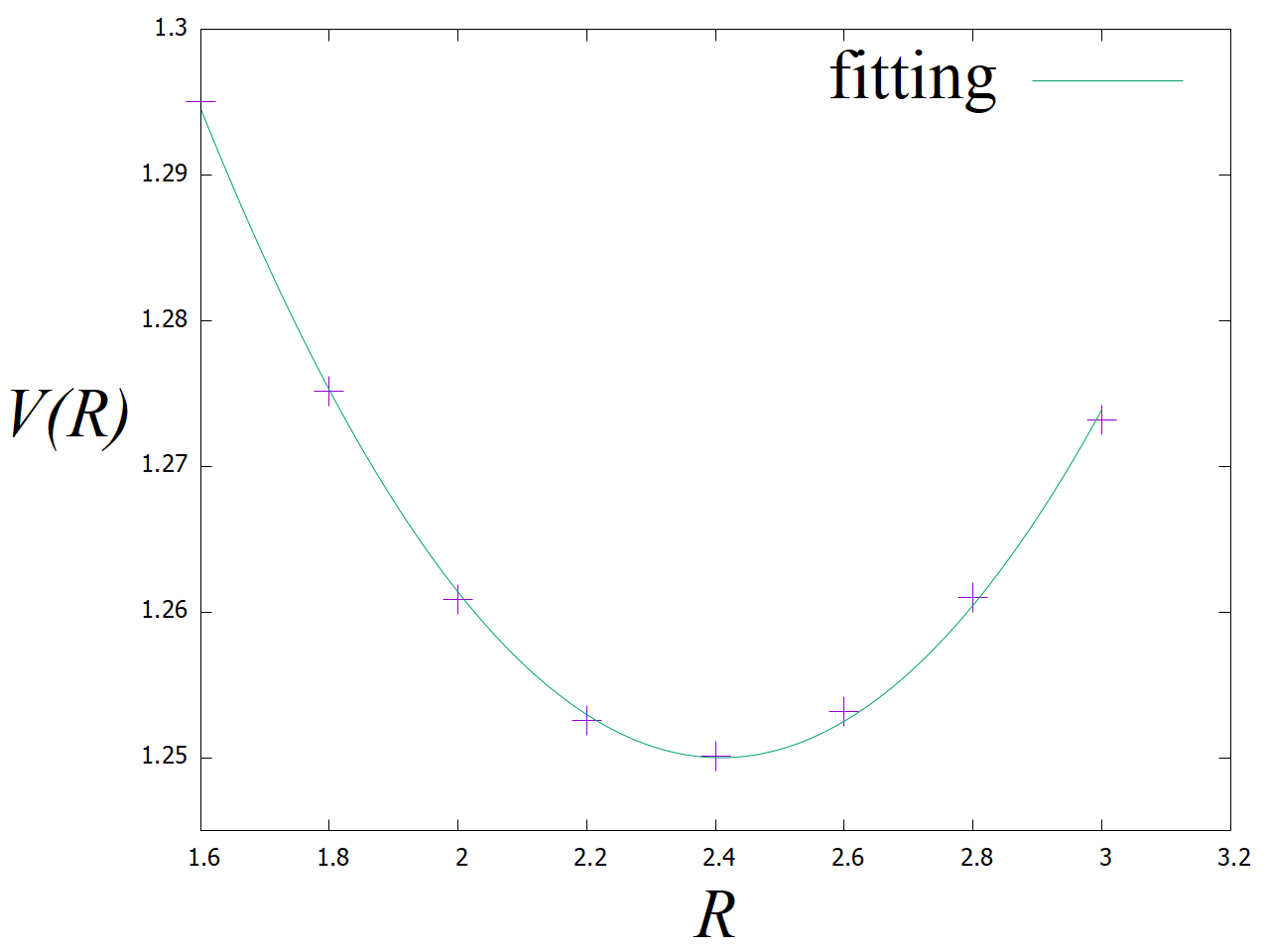}
    \end{center}
    \caption{
    Static energy $V(R)$ of the baryon with various size $R$ in the $M_{\rm KK}=1$ unit. 
    The symbol denotes numerical data of the baryon energy with different sizes. 
    The line denotes a fit curve of the potential by a quadratic function. 
    This potential has a minimum corresponding to the ground-state, 
    and its value is consistent with the previous ground-state calculation. 
    }
    \label{fig:dilatation_potentail}
\end{figure}
Figure~\ref{fig:dilatation_potentail} shows 
the static energy $V(R)$ of the baryon with various size $R$. 
The potential minimum corresponds to the ground-state baryon. 
The numerical data are well fitted 
by a quadratic function, as shown in Fig.~\ref{fig:dilatation_potentail}. 
Note that all the symbols represent the solution of holographic QCD 
under the constraint of size fixing, 
which can be regarded as a boundary condition. 

We obtain the fitting quadratic function,
\begin{eqnarray}
&& V(R) \simeq A (R-R_0)^2 + M_0, \cr 
&& ~~~~~ \label{eqn:potential fitting data} \\
&&	A \simeq 0.063, \quad 
	R_0 \simeq 2.4, \quad
	M_0 \simeq 1.25, \notag{}
\end{eqnarray}
in the $M_{\rm Kk}=1$ unit.  
The value $M_0$ of potential minimum coincides with the previous calculation of the ground-state baryon mass $M_B \simeq 1.25~M_{\rm KK}$ in Eq.~(\ref{M_B}). 

The size parameter $R=R_0$ which minimizes the static energy $V(R)$ of the baryon is found to be $R_0 \simeq 2.4$ 
in the $M_{\rm KK}=1$ unit. 
This result coincides with the ground-state result shown in Fig.~\ref{fig:phi a GS}, 
where the Higgs zero-point $\zeta_r$ locates at about $ R = R_0$. 
If there were no non-trivial gravity 
($h(w)$, $k(w)$) and no CS term, 
this size dependence would disappear and the potential $V(R)$ would become flat 
because the instanton size is originally a modulus, reflecting classical scale invariance of the Yang-Mills theory. 
In fact, this size dependence of holographic baryon mass originates from those gravitational effects and CS term. 

\subsection{Analysis of holographic baryon size}

In this subsection, we investigate  
actual size of holographic baryons 
obtained in the previous subsection 
in terms of the size parameter $R$, 
which corresponds to the instanton size. 

In our calculation, 
for each constraint of fixing the size $R$, 
we already obtain the topological density $\rho_B$ as
\begin{eqnarray}
    \rho_B &=& \rho_B(r,w;R). 
\end{eqnarray}
Using this gauge-invariant local quantity, 
we investigate the size of the holographic baryon 
for each direction of $r$ and $w$. 
Here, we define 
$\rho_B$-weighted 
average of arbitrary $(x,y,z)$-rotational symmetric variable 
$O(r,w)$ as 
\begin{eqnarray}
    \langle O \rangle_{\rho_B(R)} &\equiv& 
    \frac{\int_0^\infty drr^2 \int_{-\infty}^\infty dw
    \rho_B(r,w;R) O(r,w)}
    {\int_0^\infty drr^2 \int_{-\infty}^\infty dw
    \rho_B(r,w;R)}. ~~~~ 
\end{eqnarray}

In the ordinary four-dimensional Euclidean Yang-Mills theory, 
the Pontryagin density $\rho$ of a single instanton 
with the size parameter $R$ and its center at the origin 
is given by 
\begin{eqnarray}
    \rho = \frac{1}{16\pi^2}{\textrm tr}\left( F_{\mu\nu}\tilde{F}_{\mu\nu} \right) = \frac{6}{\pi^2}\frac{R^4}{(x_\mu^2+R^2)^4},
\end{eqnarray}
and the mean square radius weighted with $\rho$ 
is evaluated as 
\begin{eqnarray}
\langle r^2 \rangle_\rho= \frac{3}{2}R^2, \quad
\langle t^2 \rangle_\rho=\frac{1}{2}R^2
\quad {\rm (YM~instanton)},
\end{eqnarray}
or equivalently
\begin{eqnarray}
\sqrt{\frac{2}{3}\langle r^2 \rangle_\rho}= 
\sqrt{2 \langle t^2 \rangle_\rho}=R
\quad {\rm (YM~instanton)},
\label{eq:instanton_size}
\end{eqnarray}
for $r\equiv (x^2+y^2+z^2)^{1/2}$ and the Euclidean time $t$.

The holographic baryon is able to be dilatated simultaneously in both $r$ and $w$ direction, imposing the Higgs field zero-point constraints, 
and we obtain various size baryons. 
Based on the above relation, 
we consider size parameters 
of the holographic baryon 
in $r$ and $w$ directions, respectively. 
In a similar manner to Eq.~(\ref{r^2_rho_B}) in Sec.~IV, 
we define the radius weighted 
with the topological density $\rho_B$ as 
\begin{eqnarray}
    d_r(R) \equiv \sqrt{\frac{2}{3}\langle r^2 \rangle_{{\rho}_B(R)}}, \quad d_w(R) \equiv \sqrt{2\langle w^2 \rangle_{{\rho}_B(R)}}, ~~~~~~ 
\end{eqnarray}
in the direction $r$ and $w$, respectively. 
To compare with the instanton size parameter $R$, 
the factors, $2/3$ and $2$, have been introduced, considering Eq.~(\ref{eq:instanton_size}). 
In fact, the ordinary self-dual Yang-Mills instanton satisfies  
$d_r(R)=d_w(R)=R$. 

\begin{figure}[h]
    \begin{center}
    \includegraphics[width=90mm]{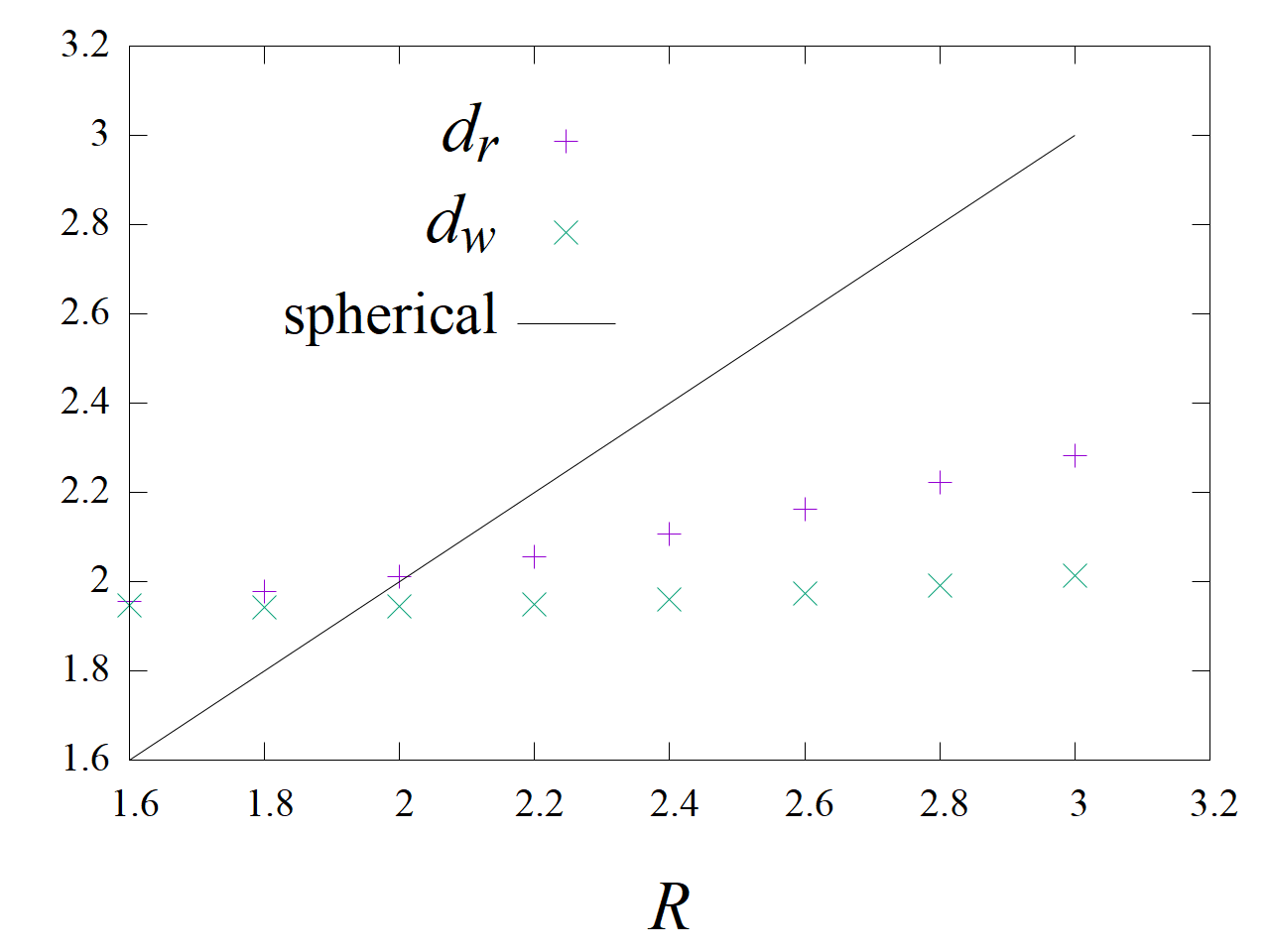}
    \end{center}
    \caption{$\rho_B$-weighted radius $d_r(R)$ and $d_w(R)$ of the holographic baryon in the direction $r$ and $w$, respectively, as the function of the size parameter $R$
    in the $M_{\rm KK}=1$ unit. 
    The solid line denotes $d=R$, 
    which is to be realized in the case of the ordinary four-dimensional YM theory  ($h(w)=k(w)=1$) without the CS term. 
    If the holographic baryon were described with the 't~Hooft solution, $d_r(R)=d_w(R)=R$ would be satisfied. 
    }
    \label{fig:shape}
\end{figure}

Figure~\ref{fig:shape} shows the $\rho_B$-weighted radius $d_r(R)$ and $d_w(R)$ as a function of $R$, 
weighted with the topological density. 

The solid line denotes $d=R$ and 
is realized in the case of the ordinary four-dimensional YM theory ($h(w)=k(w)=1$) without the CS term. 
In other words, if the holographic baryon were described with 
the 't~Hooft solution, one would find $d_{r}(R)=d_{w}(R)=R$. 

From Fig.~\ref{fig:shape}, 
$d_r$ and $d_w$ are found to be monotonically 
increasing along with $R$. 
This monotonical increase reflects 
that the change of Higgs zero-point gives 
the change of holographic baryon (instanton) size. 
Later, we investigate the dilatation mode 
by regarding the size $R$ as dynamical degree of freedom. 

Around the ground state ($R=R_0\simeq 2.4$), 
the value of $d_r(R)$ is larger than $d_w(R)$, 
i.e., $d_r(R) > d_w(R)$, 
and this means an oblate-shaped instanton 
for the holographic baryon \cite{RSR14}. 

The slope of $d_r(R)$ is larger than $d_w(R)$, 
and $d_w(R)$ is almost flat against $R$. 
Therefore, the size change of holographic baryons 
is approximately regarded as 
three-dimensional in $r$-direction 
rather than four-dimensional. 
(See Appendix C and D for 
four-dimensional size change 
using the BPS instanton.) 

These differences of the behavior for each direction 
would come from the nontrivial gravity fields, 
$h(w)$ and $k(w)$, 
because the CS term (\ref{eqn: S CS}) equally acts 
in $r$ and $w$-direction and 
does not break $(x,y,z,w)$ $O(4)$ symmetry. 
In fact, if $h(w)=k(w)=1$, $O(4)$ symmetry is exact and 
no $(r,w)$-asymmetry appears. 
Therefore, $h(w)$ and $k(w)$ are the very origin of 
$(r,w)$-asymmetry. 

The flatness of $d_w(R)$ against $R$ 
indicates that gravity fields $h(w)$ and $k(w)$ 
suppress the $w$-direction swelling. 
Approximately, one finds $d_r(R) \sim d_w(R)$, 
and then $d_r(R)$ seems to follow $d_w(R)$ 
and its soft $R$-dependence, 
which might imply that 
large deviation from the spherical shape 
is not favored energetically. 
As the result, the slope of both parameters 
become small. 

In the original Yang-Mills theory, 
the (anti)instanton appears as the (anti)self-dual solution.
In holographic QCD, owing to the presence of gravity ($h(w)$ and $k(w)$) and the CS term, 
the self-duality of the solution is explicitly broken. 

\begin{figure}[h]
    \begin{center}
    \includegraphics[width=90mm]{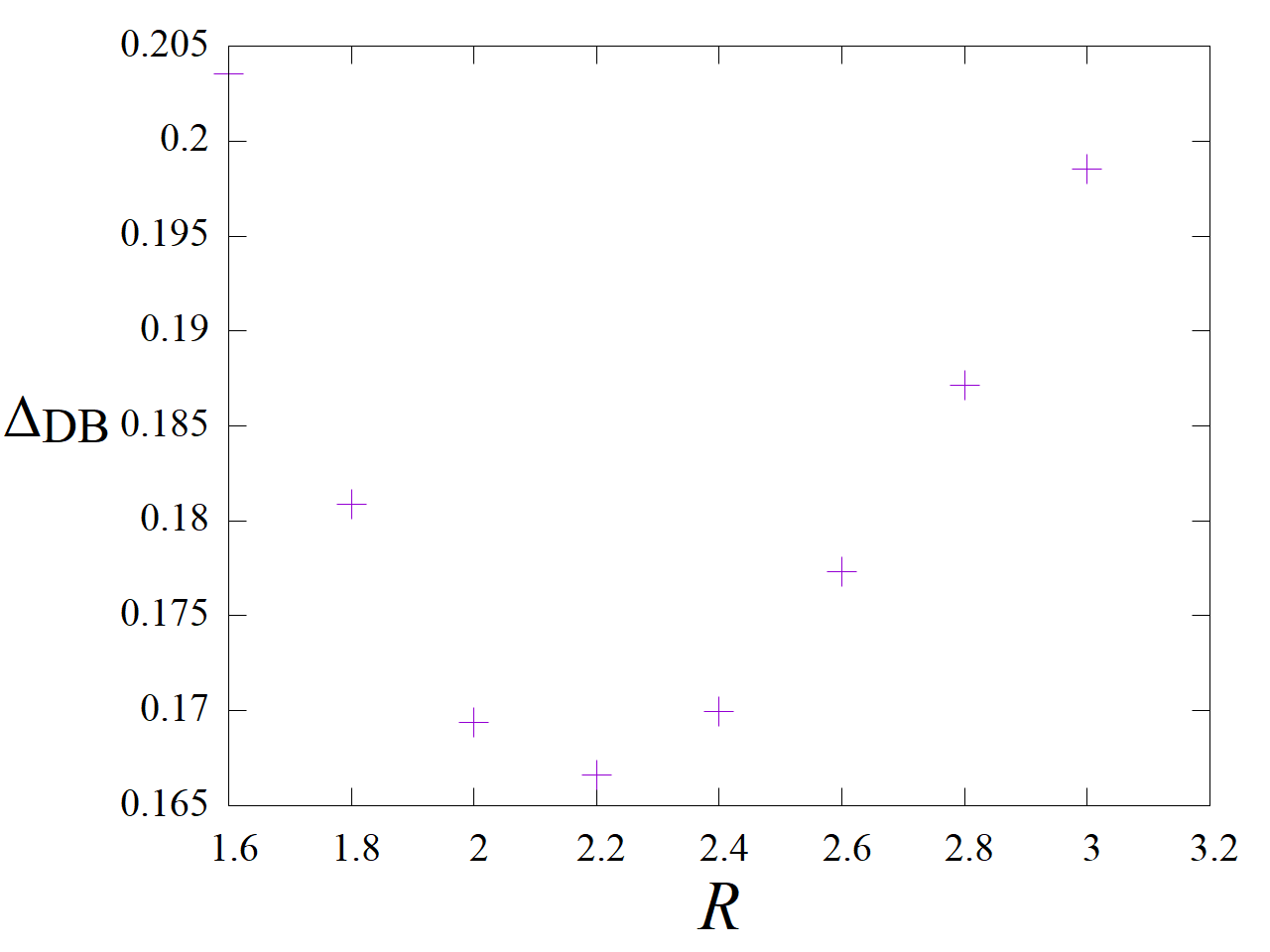}
    \end{center}
    \caption{Duality breaking parameter $\Delta_{\rm DB}$ for various baryon size $R$ in the $M_{\rm KK}$=1 unit. 
     The functional form of $\Delta_{\rm DB}(R)$ seems to be quadratic and it takes minimum at $R\simeq 2.2$. 
    }
    \label{fig:duality_breaking}
\end{figure}

Similarly for the ground-state baryon in Sec.~IV, 
we investigate self-duality breaking for 
various size holographic baryons. 
Figure~\ref{fig:duality_breaking} 
shows the self-duality breaking parameter 
$\Delta_{\rm DB}$ 
in Eq.~(\ref{eq:self-duality breaking parameter}) 
as a function of the size parameter $R$.
This quantity $\Delta_{\rm DB}$ is non-negative and 
becomes zero only in the exact self-dual case. 

One finds that the duality breaking parameter is minimized as 
$\Delta_{\rm DB} \simeq 0.17$ at $R\simeq 2.2$, which is close to  
the size $R_0 \simeq 2.4$ of the ground-state baryon. 
This value $\Delta_{\rm DB} \simeq 0.17$ seems to be small, 
and one might expect that the approximation of 
using the self-dual solution is not so bad. 

However, the duality breaking parameter $\Delta_{\rm DB}$ 
never becomes zero in holographic QCD, 
and this fact might have an important physical meaning 
for the baryon-baryon interaction as follows.

In the original four-dimensional Yang-Mills theory, 
the energy is classically bounded by BPS bound, 
and its minimum is achieved only if the configuration has self-duality.
However, holographic QCD has a non-trivial gravity and the CS term, 
and thus they distort the self-duality. 
If there were multi instantons satisfying BPS saturation, 
its action would be determined only by the topological charge, 
indicating the ``no interaction" between instantons.
Then, as an interesting possibility, 
the self-duality breaking in holographic QCD might be related to 
the baryon-baryon interaction or the nuclear force \cite{HSS09}. 

\section{Dilatation mode of a single baryon}
\label{sec:dilatation}

In the previous section, 
we investigated size dependence of the static energy $V(R)$ 
of a holographic baryon and showed that 
it seemed to be quadratic against the size $R$. 
Using this result, we numerically investigate 
time-dependent size oscillation modes, i.e., dilatation modes, 
of a single baryon in holographic QCD in this section. 

Since the instanton size $R$ is a key parameter 
to determine the baryon size in holographic QCD,
we describe the size oscillation of the holographic baryon 
by introducing time-dependence of size $R$, 
\begin{eqnarray}
    R \rightarrow R(t) = R_0 + \delta R(t), 
\end{eqnarray}
where $R_0$ denotes the size of the ground-state baryon 
and the size $R(t)$ is expected to oscillate around this $R_0$. 

Note again that the size of an instanton is originally a moduli, 
and, if there were only the Yang-Mills term in flat space $h(w)=k(w)=1$, 
its value would not affect the energy of holographic QCD 
and the dilatation mode would appear as an exact zero mode. 
In reality, the gravity effect ($h(w)$, $k(w)$) 
and CS term break this moduli property 
of the size; however, we expect that 
the dilatation mode is inherited to be a soft mode
of the ground-state baryon  
and appears as a low-lying excitation 
in the single baryon spectrum in holographic QCD. 
As a characteristic property, 
the dilatation never changes the flavor 
and rotational properties and hence 
the dilatation excitation has the same quantum number 
as the ground state. 

This dilatation differs from an ordinary $(x,y,z)$-spatial size oscillation, 
because holographic QCD has an extra dimension $w$ 
and the holographic baryon extends also its direction. 
In fact, this extra-dimensional dilatation is peculiar to holographic QCD. 

We consider time-dependent variation of $R(t)$ 
around the ground-state size $R_0$, 
which physically means a dilataion of 
the holographic baryon. 
The Lagrangian of the size variable $R(t)$ is written as 
\begin{eqnarray}
    L[R] &=& \frac{1}{2} m_R \dot{R}^2 -V(R) \cr 
    &\simeq& \frac{1}{2} m_R \dot{R}^2 - \frac{1}{2} m_R \omega^2 (R-R_0)^2 
\end{eqnarray}
up to $\mathcal{O}((R-R_0)^2)$. 
We have already calculated the potential term $V(R)$ 
and have shown it to be almost quadratic
in the previous section. 

We calculate the dilatation mode of the holographic baryon  
as a collective coordinate motion of the size $R(t)$.
Note here that, to estimate the frequency $\omega$ of the dilatational mode,  
one only has to calculate the mass parameter $m_R$.
(Of course, $m_R$ is not equal to the baryon mass $M_B$.)
Here, we use adiabatic approximation 
that time-dependence of the holographic fields 
is only through the size $R(t)$. 

\subsection{Numerical calculation}

In this subsection, 
we consider the baryon dilatation mode 
and investigate the excitation energy 
with keeping the gravity background, $h(w)$ and $k(w)$. 
We treat the size oscillation to be adiabatic 
and field motions are relatively faster than the size motion. 
Within this adiabatic treatment, 
the time dependence of holographic fields 
$\phi$ and $a_i$ is decided through only 
the baryon size $R(t)$.
Therefore, it is possible to write down field arguments symbolically as
\begin{eqnarray}
    \phi = \phi(r,w;R(t)), \quad
    a_i = a_i(r,w;R(t)) .
\end{eqnarray}
Note that there is no contribution from $\hat{A}_0$ 
in calculating the kinetic term of size variable $R(t)$, because $\hat{A}_0$ has no time-derivative 
and never accompanies $\dot{R}(t)$. 

Based on this adiabatic treatment, 
time-derivative is converted to $R$-derivative as 
\begin{eqnarray}
    \frac{d}{dt} O[R(t)] = \dot{R} \frac{d}{dR} O[R]. 
\end{eqnarray}
In our framework, 
$O[R]$ is (numerically) calculable for arbitrary $R$, 
and the $R$-derivative is easily obtained numerically,  
\begin{eqnarray}
    \frac{d}{dR} O[R] \simeq \frac{O[R+\delta R] - O[R]}{\delta R}. 
\end{eqnarray}
We have already obtained holographic configurations $\phi(R(t))$ 
with various size $R$, 
and the kinetic term of $R(t)$ is expressed as
\begin{eqnarray}
    S_{\rm kin} & = & 4 \pi \kappa \int dt\ \int_0^\infty dr\ \int_{-\infty}^\infty dw \cr 
    & & \left[ h |\partial_0\phi|^2 + h\frac{r^2}{2}(\partial_0 a_1)^2  + k\frac{r^2}{2}(\partial_0 a_2)^2 \right] \cr
    & = & 4\pi\kappa \int dt\ \int_0^\infty dr\ 
    \int_{-\infty}^\infty dw \cr
    & & \left[ h \dot{R}^2 |\partial_R\phi|^2 + h\frac{r^2}{2}\dot{R}^2(\partial_R a_1)^2 + k\frac{r^2}{2}\dot{R}^2(\partial_R a_2)^2 \right] \cr
    & = & \int dt \ \frac{1}{2} m_R \dot{R}^2 , 
\end{eqnarray}
and mass parameter $m_R$ on the size variation 
is given by
\begin{eqnarray}
    m_R & \equiv & 8\pi\kappa \int^\infty_0 dr \int^\infty_{-\infty} dw \cr & & \left[ h|\partial_R\phi|^2 + h\frac{r^2}{2}(\partial_R a_1)^2 + k\frac{r^2}{2}(\partial_R a_2)^2 \right] . 
    \label{eqn:m_R}
\end{eqnarray}
Note that CS term also contains time derivative, 
but it is first order and 
no effect from the CS term for the kinetic term of dilatation mode. 
The numerical result is found to be 
\begin{eqnarray}
    m_R \simeq 0.34~M_{\rm KK} \simeq 322~{\rm MeV}. 
\end{eqnarray}
The important point of this numerical calculation is that gravitational factors, $h(w)$ and $k(w)$,  
are exactly included.

With this value of the mass parameter $m_R$ and the quadratic fitting of 
Eq.~(\ref{eqn:potential fitting data})
for the potential $V(R)$, we obtain 
the dilatational excitation energy 
 \begin{eqnarray}
    \omega = \sqrt{\frac{2A}{m_R}} \simeq 0.61 M_\textrm{KK} \simeq 577~\textrm{MeV} 
    \label{eq:dilatational excitation energy}
\end{eqnarray}
for the holographic baryon. 

In terms of $1/N_c$ expansion, 
the mass parameter 
$m_R \propto \kappa$
in Eq.(\ref{eq:dilatational excitation energy}) 
is $O(N_c)$, 
and the potential $V(R)$ of 
the holographic baryon is $O(N_c)$, leading to 
$A = O(N_c)$. 
Then, the dilatation excitation energy $\omega = \sqrt{2A/m_R}$ is $O(N_c^0)=O(1)$ quantity.

In Appendix~\ref{appendix:Rough analytical estimation for dilation modes}, 
we also consider a rough analytical estimation for the dilatation mode, 
when using 't~Hooft instanton, i.e., 
the solution in the case of $h(w)=k(w)=1$ without the CS term. 
This rough estimate gives a larger value of 771~MeV 
for the dilatation excitation energy, 
which seems consistent with 
0.816 $M_{\rm KK} \simeq$ 774 MeV of 
the excitation energy on instanton-size fluctuation  
in the previous research \cite{HSSY07} 
using 't~Hooft instanton. 

\subsection{Discussion}

In the previous subsection, 
we numerically calculate the dilatation excitation energy, 
keeping the gravitational effect of $h(w)$ and $k(w)$. 
Note again that this background gravity is physically important 
because it is inherited from the $N_c$ D4 branes 
to express the original Yang-Mills theory. 
We have consistently performed a numerical calculation 
for both kinetic and potential terms 
to include the gravitational effect. 
We have eventually obtained 
the dilatation excitation energy of 577~MeV for 
the holographic baryon. 

This dilatation mode appears as an excited baryon with 
the same quantum number as the ground state 
since this rotationally symmetric dilatation 
never changes the quantum numbers on the flavor and rotation. 

Similar to the Skyrme soliton, 
for the description of definite spin/isospin states 
like N and $\Delta$, 
semi-classical quantization by adiabatic rotation 
\cite{HSSY07} is used for the holographic baryon. 
In this quantization process on the spin/isosipn, 
an $O(1/N_c)$ mass correction  
is added to the $O(N_c)$ baryon mass \cite{HSSY07}. 

In terms of $1/N_c$ expansion, 
the dilatation excitation energy is 
$O(N_c^0)$ as mentioned before, 
and thus the dilatation mode is more significant 
than the $O(1/N_c)$ rotational energy, 
which leads to N-$\Delta$ mass splitting. 
Furthermore, 
the correction from this rotational effect 
to the dilatation mode 
is higher order of $1/N_c$ expansion 
and then becomes negligible. 

\begin{figure}[h]
    \begin{center}
    \includegraphics[width=85mm]{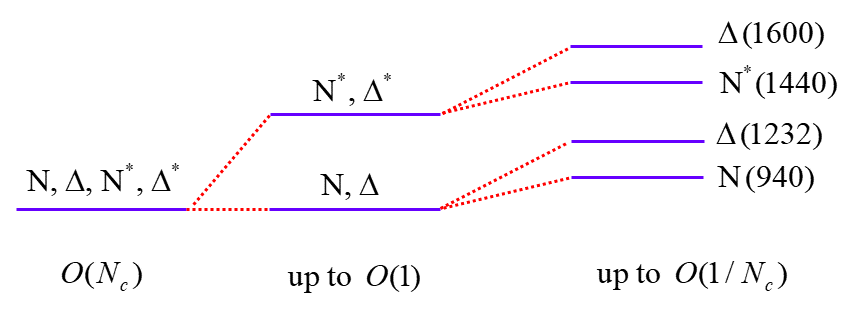}
    \end{center}
    \caption{
    Schematic figure for mass splitting 
    of the holographic baryon 
    in terms of the $1/N_c$ expansion.
    (The symbol * denotes the excitation mode.)
    The holographic baryon has 
    $O(N_c^0)$ and $O(N_c^{-1})$ splitting, 
    corresponding to dilatation and rotational effect, respectively. 
    Therefore, the order of low-lying baryon mass is to be 
    ${\rm N} < \Delta < {\rm N}^* < \Delta^*$ 
    in term of the $1/N_c$ expansion. 
    }
    \label{fig:baryon splitting}
\end{figure}

Figure~\ref{fig:baryon splitting} 
shows schematic figure for the baryon mass splitting 
order by order in the $1/N_c$ expansion. 
At the leading order of $O(N_c)$, 
all the holographic baryons degenerate. 
Up tp $O(N_c^0)$, there appears 
the mass splitting of dilatation mode. 
Up to $O(N_c^{-1})$, 
the mass splitting from rotational effect 
is added and leads to N-$\Delta$ mass splitting. 
As the result, 
the order of low-lying baryon mass is to be 
${\rm N} < \Delta < {\rm N}^* < \Delta^*$ 
in term of the $1/N_c$ expansion. 

In holographic QCD, the dilatation excitation seems to appear as the first excitation of the ground-state baryon and has the same quantum number, e.g., positive parity. 
In addition to the qualitative properties, 
the dilatation excitation energy is estimated as 577 MeV 
in holographic QCD. 
From these results on the same quantum number 
and the magnitude of the excitation energy, 
this dilatation mode of the nucleon N(940) 
would be identified as the Roper resonance N$^*$(1440),
which is positive parity and the first excitation of N(940). 
For the $\Delta$(1232), 
$\Delta$(1600) might be identified to be 
the dilatational excitation mode. 

As an interesting possibility, 
the similar dilatation mode exists for every baryon 
as a universal phenomenon for a single baryon, 
because any extended stable soliton generally has such a dilatational mode. 

The strange baryon mass is slightly larger, 
reflecting the non-zero strange quark mass; 
however, the SU(3)$_f$ flavor symmetry approximately holds. 
Here, we suppose that the mass excess of the strange quark 
is simply added to the holographic baryon, 
like the treatment of SU(3)$_f$ symmetry and its breaking 
in ordinary hadron physics \cite{GOR68}. 
Then, also for strange baryons, 
the dilatation mode would appear, 
and its excitation energy is expected to take 
a similar value as the above result of 577~MeV. 

Table \ref{tab:dilatational modes} presents the candidates of the dilational excitation in various channel of baryons, i.e., 
N, $\Delta$ and $\Lambda$, $\Sigma$, $\Sigma^*$ channel.  
\begin{table}[h]
 \centering
  \begin{tabular}{lcccc}
   \hline
   baryon & excited baryon & excitation energy & theory \\
   \hline
   N(940) & N$^*$(1440) & 500 MeV & $\omega$ \\
     & N$^*$(1710) & 770 MeV & 2$\omega$ \\
   $\Delta$(1232) & $\Delta$(1600) & 368 MeV & $\omega$ \\
    & $\Delta$(1920) & 688 MeV & $2\omega$ \\
   $\Lambda$(1116) & $\Lambda$(1600) & 484 MeV & $\omega$ \\
   $\Sigma$(1193) & $\Sigma$(1660) & 467 MeV & $\omega$ \\
   $\Sigma^*$(1385) & $\Sigma^*$(1780) & 395 MeV & $\omega$ \\
   \hline
  \end{tabular}
  \caption{
  Experimental candidates of the dilatational excitation mode in various baryon channel \cite{PDG}. 
  Each ground-state baryon has the excitation with the same quantum number. 
  Here, first-excited baryons are mainly listed, 
  such as N$^*$(1440) and $\Delta$(1600). 
  For each channel, the excitation energy seems 
  to take a consistent value with the theoretical one, 
  $\omega \simeq 577~{\rm MeV}$. Here, 
  N$^*(1710)$ and $\Delta(1920)$ are identified to the second excitation mode. 
  }
  \label{tab:dilatational modes}
\end{table}

For each channel with the same quantum number, 
we find that the first excitation energy 
seems to be a consistent similar value 
to the dilatational one, 
$\omega \simeq 577~{\rm MeV}$, 
theoretically obtained above. 

Also for multi-strange baryons 
in $\Xi$, $\Xi^*$ and $\Omega$ channel, 
we theoretically predict the dilatation modes 
with the excitation energy of about 577~MeV. 
However, 
the dilatational excited baryons for 
$\Xi(1320)$, $\Xi^*(1530)$, 
and $\Omega(1673)$ are not yet observed experimentally 
because spin-parity information 
is not yet confirmed for their excited baryons. 
In this respect, further experimental 
analyses are much desired for excited baryons in terms of 
the dilatational mode in each baryon channel. 

\section{Summary and Concluding Remarks}

We have investigated a baryon and its dilatation modes 
in holographic QCD based on the Sakai-Sugimoto model, 
which is constructed with $N_c$ D4 branes and $N_f$ D8/$\bar{\rm D8}$ branes in the superstring theory. 
This theory is expressed as 
a 1+4 dimensional U($N_f$) gauge theory 
in the flavor space.

We have adopted a generalized version of the Witten Ansatz for spatially rotational symmetric systems and have reduced 
1+4 dimensional holographic QCD into a 1+2 dimensional Abelian Higgs theory in a curved space. 
In this formulation, a four-dimensional instanton 
corresponding to a baryon is converted to 
a two-dimensional Abrikosov vortex. 
We have numerically calculated the baryon solution of 
holographic QCD using a fine and large-size lattice 
with spacing of 0.04~fm and size of 10~fm.

Using the relation between the baryon size and 
the zero-point location of the Higgs field in the Witten Ansatz,  
we have theoretically changed the size of holographic baryons  
and have investigated its properties, such as the energy and 
self-duality breaking, as function of the size parameter.
Here, each configuration is a solution of EOM of holographic QCD 
under the constraint of fixing Higgs zero-point. 

As time-dependent size-oscillation modes, 
we have investigated the dilatation modes of a baryon 
and have found that 
such a dilatational mode takes 
the excitation energy of 577~MeV. 
Since the dilatation does not change 
the quantum number including the parity, 
we have identified this dilatation mode for the nucleon N(940) 
as the Roper resonance N$^*$(1440).
We have conjectured that any baryons are expected 
to have such a dilatational excitation universally, 
and their excitation energy would be similar.

In this respect, further experimental analyses are much 
desired for excited baryons in terms of 
the dilatational mode in each baryon channel. 
In particular, the dilatational excited baryons for 
$\Xi(1320)$,  $\Xi^*(1530)$, and  $\Omega(1673)$ have not yet been observed  experimentally 
because spin-parity information is not yet confirmed for their excited baryons. 

As a caution, the calculated values presented in this paper 
are to be regarded as semi-quantitative estimates 
in an idealized case of large $N_c$ in the chiral limit, 
and they have some deviation from experimental values in the real world. 
In the following, we mention some quantitative limitations 
and cautions on the approach used in this study. 

This framework of holographic QCD has infrared equivalence 
with massless QCD and gives a useful analytical 
nonperturbative method to analyze QCD, 
which is a main reason to adopt holographic QCD in this study for baryons. 
On the quantitative accuracy, however, 
this framework includes some limitations. 
As an important caution, 
the present calculation is based on the 
$1/N_c$ and $1/\lambda$ expansion, 
and the starting holographic action is up to 
the $1/N_c$-leading and $1/\lambda$ sub-leading order. 
Therefore, for more accurate estimation, 
it is desirable to check the contribution from 
$1/N_c$ or $1/\lambda$ higher order terms. 
However, it is extremely difficult to extract 
the next order of $1/N_c$ and $1/\lambda$ expansions 
in holographic QCD. 

As a higher-order correction, 
there is an $O(1/N_c)$ rotational effect of 
the hedgehog configuration, 
which is used for semi-classical analysis 
of the Skyrmion investigation \cite{ANW83}. 
It is relatively $1/N_c^2$ smaller, 
compared with leading order $O(N_c)$ of the baryon mass, 
although it is necessary for the $N-\Delta$ splitting. 
Similarly, for the dilatational excitation, 
this correction appears as relatively $1/N_c^2$-smaller order, 
and therefore we have ignored this higher order in this paper. 
However, this higher-order correction might be desired 
to reproduce real experimental data. 

In addition, we have assumed spatially spherical shape 
of a baryon to apply the Witten Ansatz 
and adiabatic treatment for the dilatation dynamics.  
For more precise analysis of baryons, 
more sophisticated treatments might be desired. 
However, to go beyond the spherical symmetric solution, 
the Witten Ansatz is no more applicable, 
and then one has to deal with the four-dimensional analysis even for static baryons. 
To go beyond adiabatic approach 
is also a difficult problem widely appeared 
in theoretical physics, and 
one has to handle complicated local oscillation 
of all the holographic fields in the present case. 

We have used holographic QCD based on $N_f = 2$ 
in the chiral limit. 
Further extension including strangeness is 
an interesting subject in hadron physics, 
which can be done with $N_f=3$ holographic QCD \cite{MS17}. 
In the real world, u and d-quarks have 
small finite current mass of 2-5~MeV 
and s-quarks have current mass of about 93~MeV \cite{PDG}, 
and the non-zero quark mass explicitly breaks the chiral symmetry. 
In holographic QCD, however, it is difficult to introduce 
the finite quark mass or explicit chiral-symmetry breaking, 
and further theoretical development is required 
for quantitative argument of hyperons. 

We have mainly presented 
the excitation energy for the dilatational mode of baryons,
which appears in the same quantum numbers. 
It is desired for theoretical progress 
to show how to distinguish 
the dilatational mode from other excitation 
experimentally. 
To this end, we have to find out 
unique behavior for the dilatational mode, 
which is a future subject. 
The finding of such a peculiar quantity will 
lead to a better understanding of baryon spectra. 

\begin{acknowledgements}
H.S. is supported in part by the Grants-in-Aid for
Scientific Research [19K03869] from Japan Society 
for the Promotion of Science.
\end{acknowledgements}

\newpage

\appendix

\section{Lattice formalism of holographic QCD}

In Appendix~A, 
we show our lattice formalism for holographic QCD 
in the Witten Ansatz, 
by introducing a sizable finite lattice with a small spacing $a$. 

Let us begin with the static energy $E_{\rm 5YM}$ 
of the Yang-Mills term in the Witten Ansatz. 
For the explanation, we here repeat Eq.~(\ref{eqn:E_5YM}),  
\begin{eqnarray}
E_{\rm 5YM} &=& 4\pi\kappa \int^{\infty}_{0} dr \int^{\infty}_{-\infty} dw \biggl[ h(w) |D_1\phi|^2 + k(w) |D_2\phi|^2  \cr 
& & + \frac{h(w)}{2r^2}\{1-|\phi|^2\}^2 + \frac{r^2}{2} k(w) f_{12}^2 \biggr], 
\label{eq:E_5YM_appendix}
\end{eqnarray}
of which the field variables $\phi$ and $a_\mu$ 
depend on the two-dimensional spatial coordinate $(r,w)$.

For the U(1) gauge variable $a_\mu$ with $\mu=r,w$, 
we define the link variable 
\begin{eqnarray}
U_\mu(s) \equiv \exp\{ia~a_\mu(s+\frac{\hat{\mu}}{2})\} \in {\rm U(1)}
\end{eqnarray}
at the site $s=(s_r,s_w)$ on the two-dimensional lattice.  
Here, $\hat{\mu}$ denotes the $\mu$-directed vector 
with the length of $a$.

On the lattice with a small spacing $a$, 
one finds for the U(1) covariant derivative $D_\mu\phi$ as 
\begin{eqnarray}
& & -\{ \phi^\dagger(s)U_\mu(s)\phi(s+\hat{\mu}) + \phi^\dagger(s+\hat{\mu})U_\mu(s)^\dagger\phi(s) \} \cr
& & + |\phi(s)|^2 + |\phi(s+\hat{\mu})|^2 \cr
& = & - ( \phi-\frac{a}{2}\partial_\mu\phi )^\dagger (1+iaa_\mu-\frac{1}{2}a^2a_\mu^2)(\phi+\frac{a}{2}\partial_\mu\phi)\cr
& & - ( \phi+\frac{a}{2}\partial_\mu\phi )^\dagger (1-iaa_\mu-\frac{1}{2}l^2a_\mu^2)(\phi-\frac{a}{2}\partial_\mu\phi) \cr
& & + 2|\phi|^2 + \frac{a^2}{2}|\partial_\mu\phi|^2 + \mathcal{O}(a^3) \cr
& = & a^2\phi^\dagger(\stackrel{\leftarrow}{\partial_\mu}-ia_\mu)(\partial_\mu+ia_\mu)\phi + \mathcal{O}(a^3) \cr 
& = & a^2 |D_\mu\phi|^2 + \mathcal{O}(a^3),
\end{eqnarray}
where all omitted arguments are $s+\hat{\mu}/2$. 
The field strength $f_{12}$ is expressed with 
the U(1) plaquette variable,  
\begin{eqnarray}
    \square_{12}(s) &\equiv& U_1(s)U_2(s+\hat{1})U_1^*(s+\hat{2})U_2^*(s) \in {\rm U(1)}~~~~ \\
    f_{12}^2(s) &=& \frac{1}{a^2}[1 - {\rm Re} \{\square_{12}(s) \}].
\end{eqnarray}

The additional U(1) energy $E^{\rm U(1)}$ 
in Eq.~(\ref{eq:E^U(1)rw}) is expressed by 
the coupling of the original U(1) gauge field $\hat A_0$ 
and the topological density $\rho_B$. 
Here, the temporal component $\hat A_0(r,w)$ 
is treated as a (spatially) site-variable 
on the $(r,w)$ lattice.
On the lattice, 
the baryon density is expressed as 
\begin{eqnarray}
\rho_B  & = & \frac{1}{8\pi^2r^2} [-i\epsilon_{ij}(D_i\phi)^*D_j\phi + \epsilon_{ij}\partial_ia_j(1-|\phi|^2) ] \cr
& = & \frac{1}{8\pi^2r^2} [ 2 {\rm Im} \{(D_1\phi)^*D_2\phi\} + f_{12}(1-|\phi|^2) ]~~~
\end{eqnarray}
with 
\begin{eqnarray}
    f_{12} (1-|\phi|^2) &=& \frac{1}{a^2} {\rm Im} \ \square_{12} \times (1-|\phi|^2)
\end{eqnarray}
and
\begin{eqnarray}
&& (D_1\phi)^*D_2\phi \cr
    &=& \frac{1}{4a^2} \{ U_1^*(s)\phi^*(s+\hat{1}) - U_1(s-\hat{1})\phi^*(s-\hat{1}) \} \cr
    & & \times \{ U_2(s)\phi(s+\hat{2}) - U_2^*(s-\hat{2})\phi(s-\hat{2}) \} \cr 
    &=& \frac{1}{4a^2} \left\{ \phi^*(s+\hat{1})U_1^*(s)U_2(s)\phi(s+\hat{2}) \right. \cr
    & & \left. - \phi^*(s-\hat{1})U_1(s-\hat{1})U_2(s)\phi(s+\hat{2}) \right. \cr
    & & \left. - \phi^*(s+\hat{1})U_1^*(s)U_2^*(s-\hat{2})\phi(s-\hat{2}) \right. \cr
    & & \left. + \phi^*(s-\hat{1})U_1(s-\hat{1})U_2^*(s-\hat{2})\phi(s-\hat{2}) \right\}. \cr
    & & 
\end{eqnarray}
To reduce the discretization error, 
we have used the above form for $(D_1\phi)^*D_2\phi$, 
and its lattice formalism is symbolically written as
\begin{equation}
4a^2 (D_1\phi)^*D_2\phi = 
\vcenter{\hbox{\includegraphics[width=40mm]{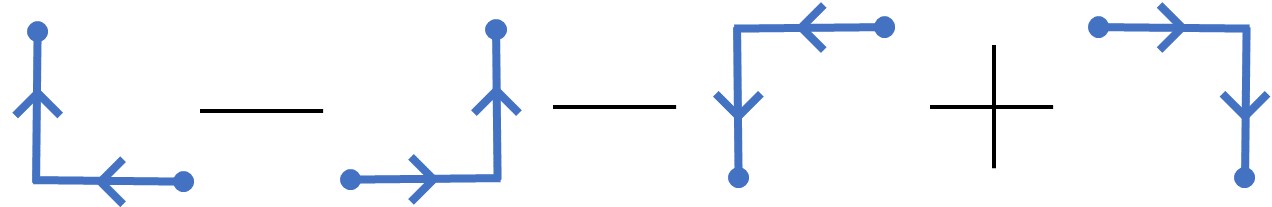}}},
\end{equation}

\ 

\noindent
where the horizontal an vertical arrows represent 
$U_1$ and $U_2$, respectively, 
and the dots represent $\phi$. 
Thus, we define the topological density $\rho_B(r,w)$ 
and formulate the U(1) energy 
$E^{\rm U(1)}[\rho_B(r,w), \hat A_0(r,w)]$
in Eq.~(\ref{eq:E^U(1)rw}) on the $(r,w)$ lattice.

In this way, the total energy $E$ in holographic QCD 
is expressed as $E[\phi(s), \vec U(s), \hat A_0(s)]$, i.e., 
a function of $\phi(s)=\phi_1(s)+i\phi_2(s) \in {\bf C}$, $\vec U(s)\equiv (U_1(s), U_2(s))$ and $\hat A_0(s)$ at the spatial site $s=(s_r,s_w)$.
To find the solution of holographic QCD, 
we minimize the total energy $E$ 
by iterative improvement on 
$\phi(s)$, $\vec U(s)$ and $\hat A_0(s)$.
Looking at a specific site $s_0$, 
we consider only one variable $\phi(s_0)$,  
with fixing all other variables.
By taking variation of $\phi(s_0)$, 
we minimize the total energy $E$. 
Next, we consider only one link-variable $\vec U(s_0)$,  
with fixing all other variables, 
and take its variation to minimize $E$. 
Similarly, considering only one site-variable $\hat{A}_0(s_0)$,  
with fixing all other variables, 
we take its variation to minimize $E$. 
On each site on the lattice, we repeat the above process  
and update $\phi(s)$, $\vec U(s)$ and $\hat A_0(s)$.
We iterate this sweep procedure many times 
so as to minimize the total energy $E$ in holographic QCD, 
and the solution is eventually obtained.

\section{Other expression of U(1)-part energy}

For the U(1) sector, we have mainly used 
Eq.~(\ref{eq:E^U(1)rw}) for 
the numerical calculation of $E^{\rm U(1)}$. 
However, there is another useful 
expression for the energy $E^{\rm U(1)}$ 
of the U(1) sector without $\hat A_0$,  
and we introduce this form in Appendix B.

By solving $\hat A_0$ 
using Eq.~(\ref{eq:U(1)FE}),
the energy (\ref{eq:E^U(1)}) becomes 
\begin{eqnarray}
E^{\rm U(1)} &=&\frac{N_c^2}{8}\int d^3 x dw~ \rho_B K^{-1} \rho_B
\nonumber \\
&=&\frac{N_c^2}{8}\int d^3 x dw \int d^3 x' dw' \cr
&&\rho_B(\vec x,w) K^{-1}(\vec x,w; \vec x',w') \rho_B(\vec x',w').
\end{eqnarray}

Since the kernel $K$ and topological density $\rho_B$ are 
SO(3) rotationally symmetric, the additional energy can be  expressed 
only with the $(r,w)$-coordinates:
\begin{eqnarray}
E^{\rm U(1)} &=& 2\pi^2 N_c^2 \int_0^\infty dr \int_{-\infty}^\infty dw 
\int_0^\infty dr' \int_{-\infty}^\infty dw'
\nonumber  \\ 
&& \tilde \rho_B(r,w) \tilde K^{-1}(r,w; r',w') \tilde \rho_B(r',w'),
\label{eqn:U(1) energy}
\end{eqnarray}
using $\tilde \rho_B(r,w) \equiv r^{2}\rho_B(r,w)$ and 
the hermite kernel $\tilde K \equiv 4\pi r^2 K$ in $(r,w)$-space. 

On the lattice, the kernel $\tilde{K}$ in Eq.~(\ref{eq:kernel_rw})
is transformed into a differential form and is expressed as 
a matrix $\tilde{K}_L(r, w; r', w')$. 
Then, the kernel inverse $\tilde{K}_L^{-1}(r, w; r', w')$ is numerically obtained 
by taking the inverse matrix of the kernel $\tilde{K}_L (r, w; r', w')$. 
As a technical caution, the kernel $\tilde{K}_L (r, w; r', w')$ 
has a translational zero mode, reflecting the derivative form of $\tilde{K}$. 
Since this translational zero mode does not affect the total energy, 
the inverse of  $\tilde{K}$ has to be taken 
in the space except the spurious zero mode.
In fact, using an orthogonal matrix $O$, the kernel $\tilde{K}_L$ is diagonalized as 
\begin{eqnarray}
    \tilde{K}_L = O \ {\rm diag}(0,\lambda_1,\lambda_2,\cdots) \ O^T
\end{eqnarray}
with non-zero eigenvalues $\lambda_n$. 
Then, we define its inverse $\tilde{K}^{-1}_L$ to be 
\begin{eqnarray}
    \tilde{K}_L^{-1} = O \ {\rm diag}(0,\lambda_1^{-1},\lambda_2^{-1},\cdots) \ O^T, 
\end{eqnarray}
which is equivalent to the appropriate removal of 
the spurious translational zero mode.
Using this kernel inverse $\tilde{K}_L^{-1}$ and the baryon density $\rho_B$, 
the energy $E^{\rm U(1)}$ of the CS term is calculated as 
$\rho_B \tilde{K}_L^{-1} \rho_B$ on the lattice. 

Thus, for the numerical calculation 
of $E_{\rm U(1)}$, there are two different methods:
one is to update the holographic fields 
$\phi(r,w)$, $\vec a(r,w)$ and $\hat A_0(r,w)$ 
on the lattice based on Eq.~(\ref{eq:E^U(1)rw});
the other is to update only 
$\phi(r,w)$ and $\vec a(r,w)$ using  
Eq.~(\ref{eqn:U(1) energy}).
We have confirmed that both methods give 
the same numerical results for holographic baryons.

\section{Holographic baryon using 
self-dual BPS instanton}

\begin{figure}[h]
    \begin{center}
    \includegraphics[width=90mm]{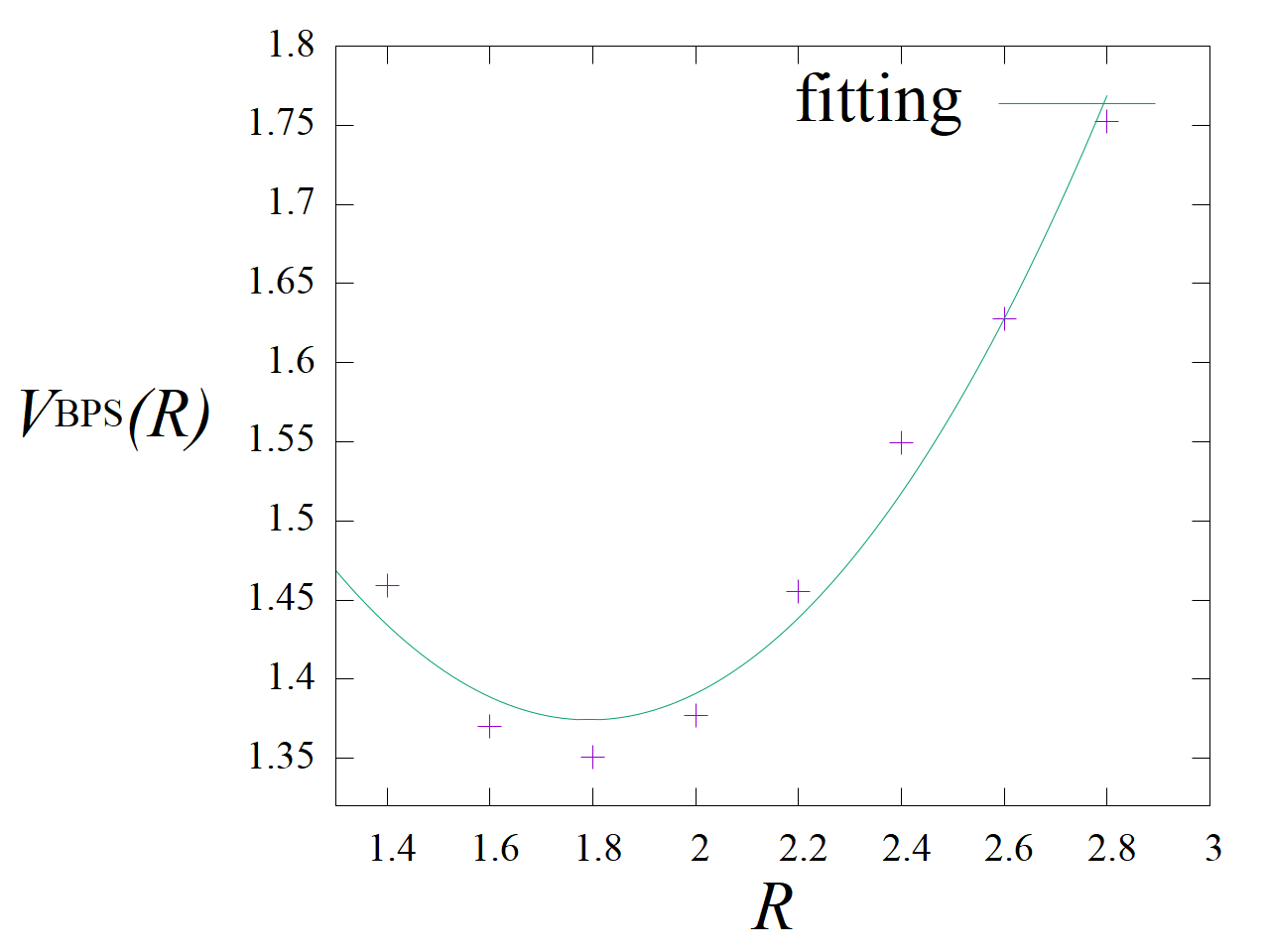}
    \end{center}
    \caption{Static baryon energy $V^{\rm BPS}(R)$ calculated from the BPS configurations in Eq.~(\ref{eqn:'t hooft soltion HQCD}). 
    The solid line is a fitting curve of quadratic function. 
    }
    \label{fig:potential BPS}
\end{figure}

In Appendix~C, 
we investigate the holographic baryon  
when the self-dual BPS instanton 
in Eq.~(\ref{eqn:'t hooft soltion HQCD})
is used.
The self-dual BPS instanton is 
the 't~Hooft solution of the ordinary Yang-Mills theory,  
and hence, to be strict, this usage 
is justified in the case of 
flat space $h(w)=k(w)=1$ and ignoring the CS term. 
Substituting the BPS instanton with the size $R$ 
into the total energy $E$ in Eq.~(\ref{eqn:total energy})
including the CS term and the background gravity, 
$k(w)=1+w^2$ and $h(w)=k(w)^{-1/3}$, 
one obtains 
the static baryon energy $E(R)=V^{\rm BPS}(R)$ 
as the function of the instanton size $R$,  
and its minimum gives 
an approximate ground-state holographic baryon, 
which satisfies 
$M_B^{\rm BPS} \simeq 1.35~M_{\rm KK} \simeq 1.28~{\rm GeV}$ 
and  
$\sqrt{\langle r^2\rangle_{\rho_B}^{\rm BPS}} 
=\sqrt{\frac{3}{2}} R_0^{\rm BPS} \simeq 2.2~M_{\rm KK}^{-1}\simeq 0.46~{\rm fm}$, 
as shown in Fig.~\ref{fig:potential BPS}.
Thus, when the self-dual BPS instanton is used, 
the holographic ground-state baryon has 
larger mass and smaller size \cite{HSSY07} 
than the true solution numerically obtained in Sec.~\ref{sec:vortex baryon}.

\begin{figure}[h]
    \begin{center}
    \includegraphics[width=90mm]{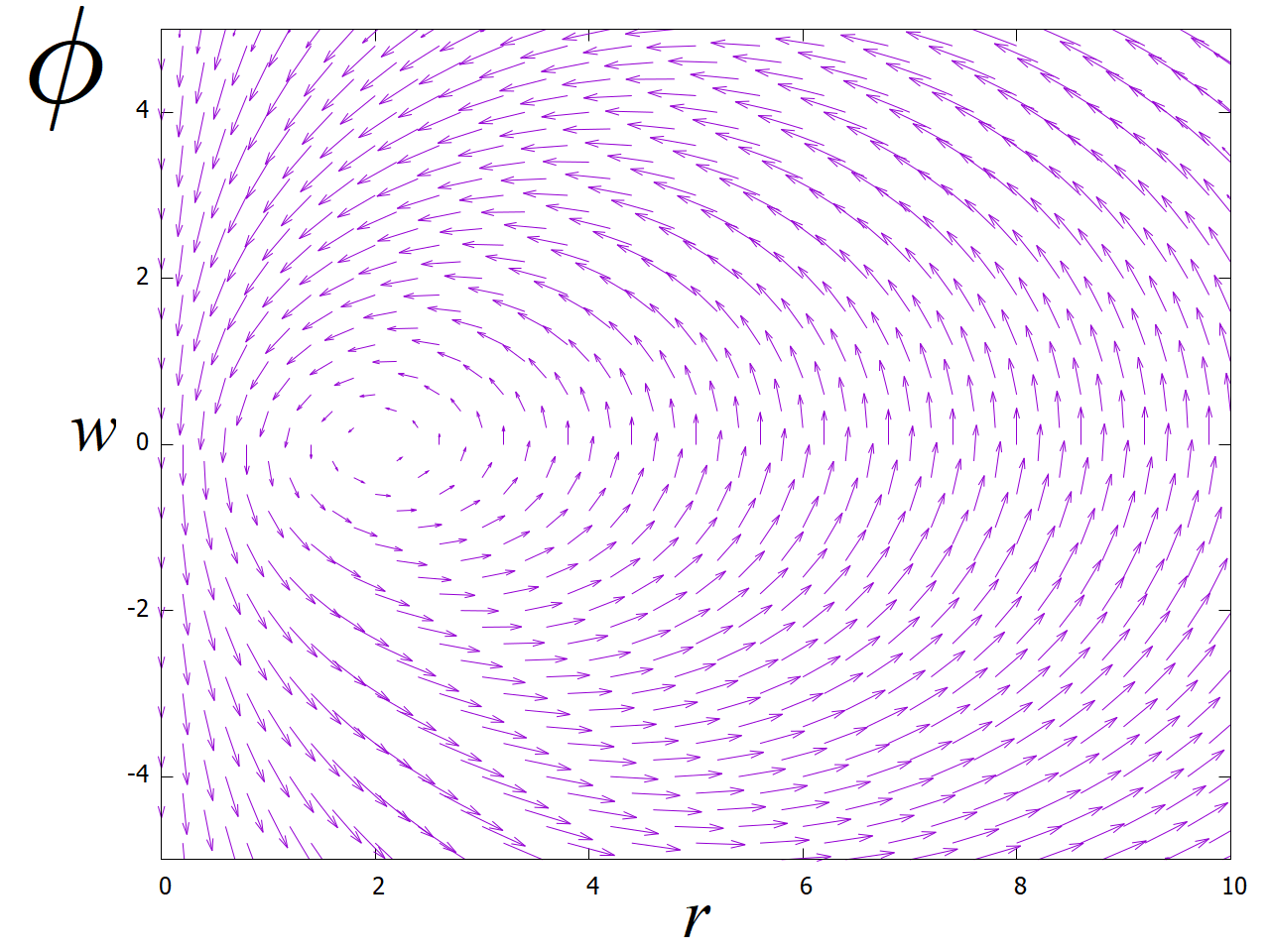}
    \includegraphics[width=90mm]{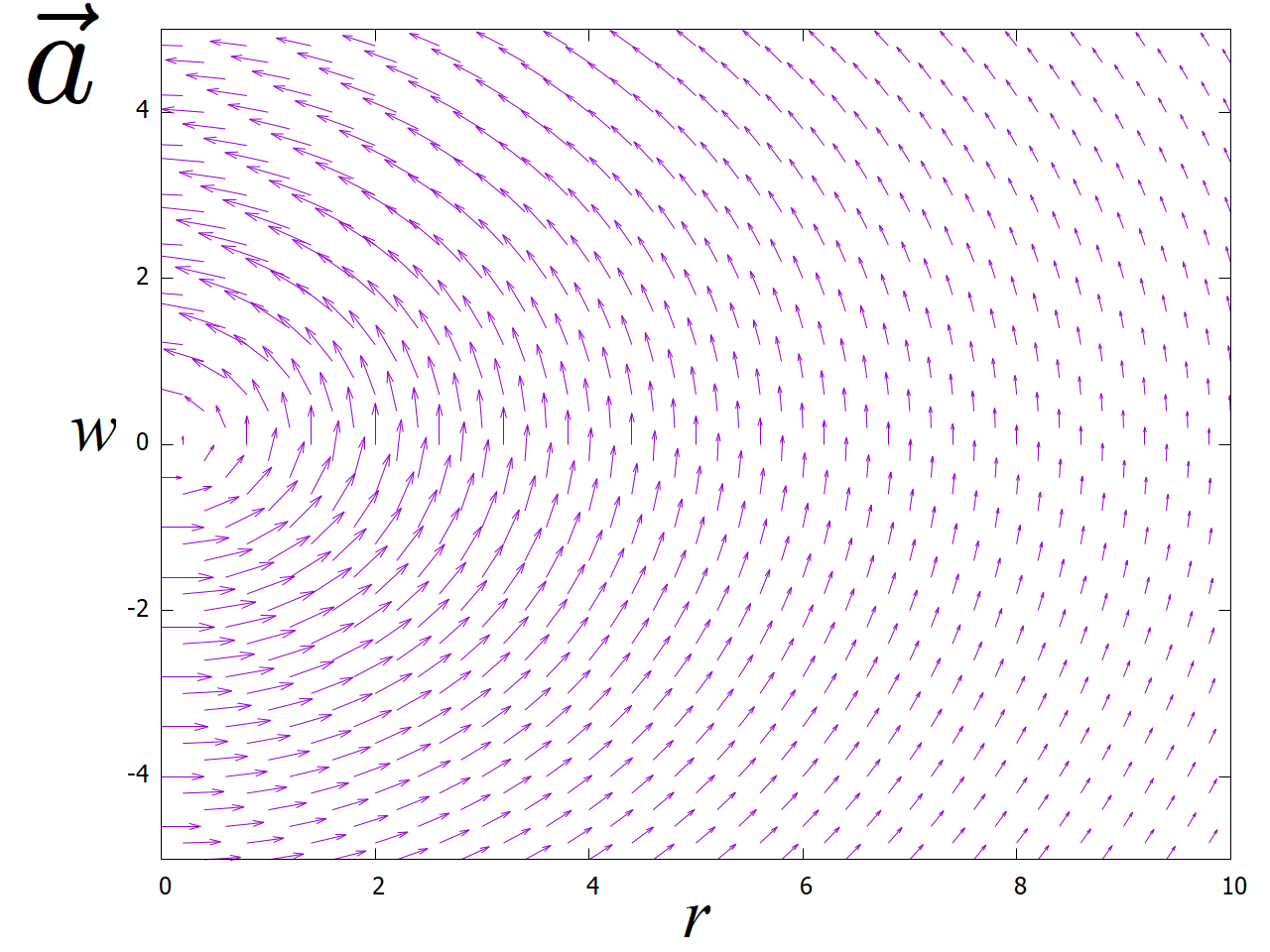}
    \end{center}
    \caption{
    The Higgs field $\phi(r,w)$ (upper) 
    and the Abelian gauge field $a(r,w) = (a_1,a_2)$ (lower) 
    for the BPS instanton in the Landau gauge in the $M_{\rm KK}=1$ unit. 
    For iterative improvement in the main sections, 
    this configuration is used as the starting point,   
    i.e., the initial configuration of the iteration. 
    }
    \label{fig:phi a no iteration}
\end{figure}

\begin{figure}[h]
    \begin{center}
    \includegraphics[width=90mm]{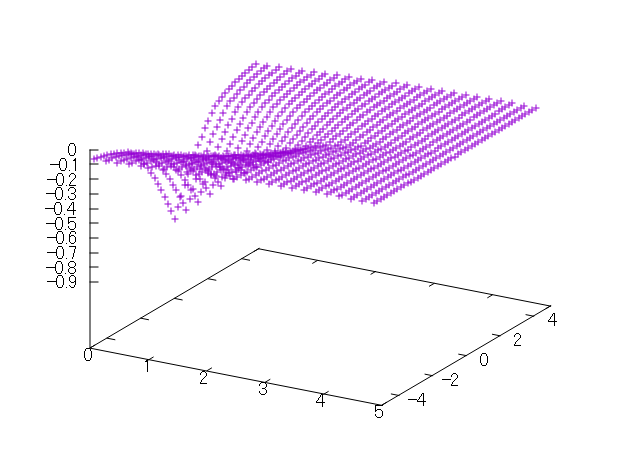}
    \end{center}
    \caption{The U(1) gauge field 
    $\hat{A}_0$ in the case of BPS instanton 
    in the $M_{\rm KK}=1$ unit. 
    This is obtained by solving Eq.~(\ref{eq:U(1)FE}) 
    for the self-dual 't~Hooft instanton. 
    }
    \label{fig:aU1 no iteration}
\end{figure}

\begin{figure}[h]
    \begin{center}
    \includegraphics[width=90mm]{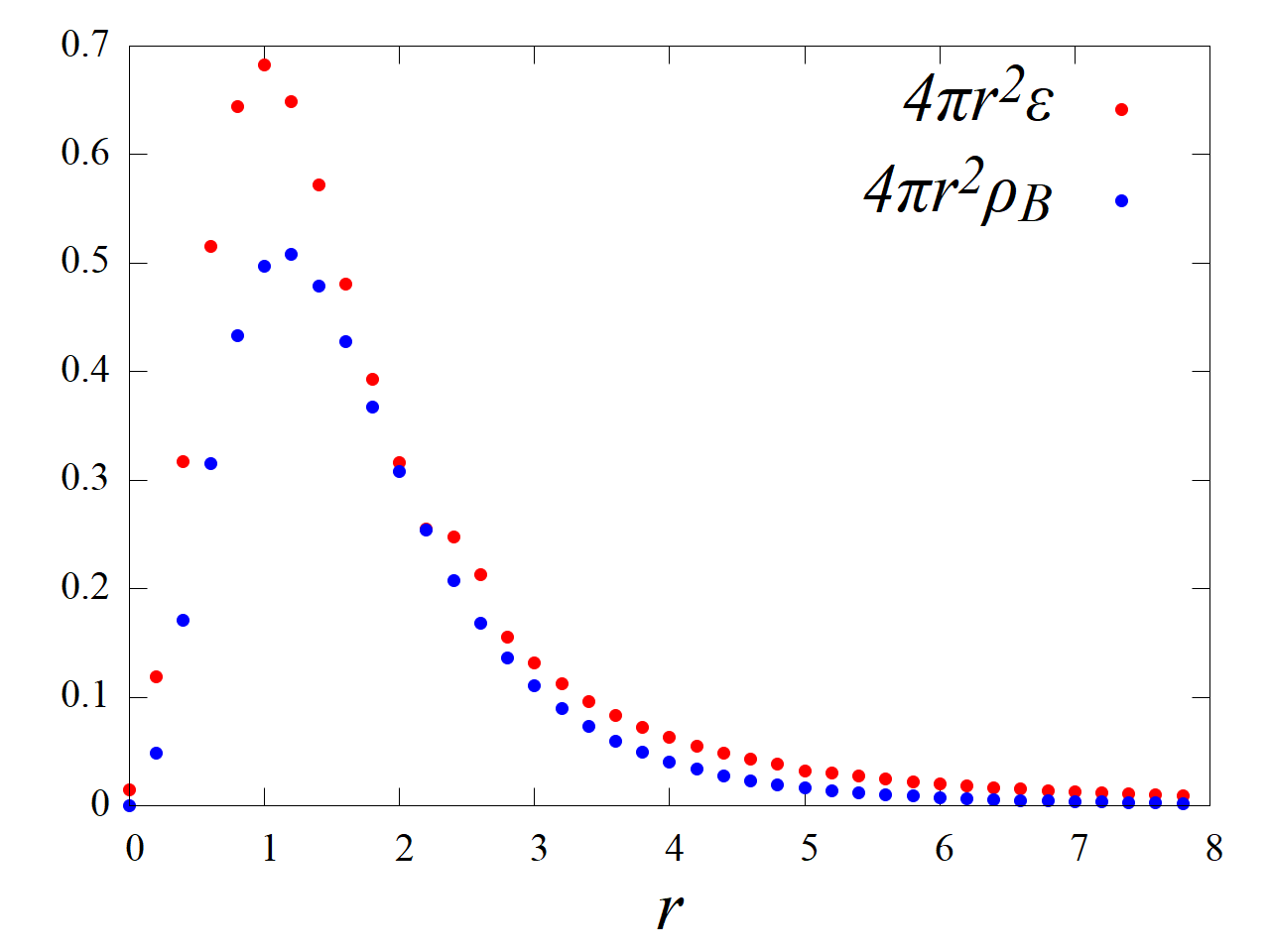}
    \end{center}
    \caption{
    $r^2$-multiplied 
    topological and energy densities, 
     $4\pi r^2\rho_B(r)$ and $4\pi r^2{\cal E}(r)$, 
    in the case of the BPS instanton configuration
    in the $M_{\rm KK}=1$ unit. 
    }
    \label{fig:e t no iteration}
\end{figure}

For the approximate ground-state holographic baryon, 
the corresponding Higgs field $\phi$ and gauge field $\vec{a}$ are shown in Fig.~\ref{fig:phi a no iteration}. 
These fields give a topological density $\rho_B$, and 
$\hat{A}_0$ is obtained by solving EOM~(\ref{eq:U(1)FE}), 
as shown in Fig.~\ref{fig:aU1 no iteration}.
Figure~\ref{fig:e t no iteration} shows 
the $r^2$-multiplied topological and the energy densities.

The static baryon energy $V^{\rm BPS}(R)$ 
for the BPS configuration is shown in Fig.~\ref{fig:potential BPS}, and 
it is approximately fit with a quadratic function,  
\begin{align}
    &&    V^{\rm BPS}(R) \simeq A^{\rm BPS} ( R - R_0^{\rm BPS} )^2 + M^{\rm BPS}, \cr 
    && \label{eqn:BPS potential data} \\
    &&    A^{\rm BPS} \simeq 0.39,~ R_0^{\rm BPS} \simeq 1.8,~ M_0^{\rm BPS} \simeq 1.4, \notag{}  
\end{align}
in the $M_{\rm KK}=1$ unit.

Thus, when the self-dual BPS instanton is used, 
the holographic baryon has 
larger mass and smaller size \cite{HSSY07} 
than the true solution numerically obtained in Sec.~\ref{sec:vortex baryon}.

\section{Rough analytical estimation for dilation modes
using self-daul BPS instanton}
\label{appendix:Rough analytical estimation for dilation modes}

In Appendix~D, we consider 
a rough analytical estimation of the dilation mode
using the self-dual 't~Hooft BPS instanton,  
which is justified in the flat space $h(w)=k(w)=1$ 
and without the CS term. 

In this case, 
when the flat space approximation $h(w)=k(w)=1$ is used, 
the kinetic term $T$ of the size variable $R(t)$ is analytically expressed as 
\begin{eqnarray}
	T & \simeq & \kappa \int d^3x\ dw \textrm{tr}\left[ F_{0M}^2 \right] \notag{} \\
	& \simeq & \kappa \int d^3x\ dw \frac{24x^2}{(x^2+R_0^2)^4} R_0^2 \dot{R}^2 \notag{} \\
	& = & 48\pi^2\kappa \int dr \frac{r^5}{(r^2+R_0^2)^4} R_0^2 \dot{R}^2 \notag{} \\
	& = & 8\pi^2\kappa \ \dot{R}^2 = \frac12 m_R \dot{R}^2, 
\end{eqnarray}
where $\dot{R}$ means $\partial_t R$. 
Here, $h(w)=k(w)=1$ is used in the first line, 
and the instanton size $R_0$ appearing in the middle 
does not affect the result. 
In this way, the mass parameter $m_R$ 
on the size variation is estimated as 
\begin{eqnarray}
 m_R & = & 16\pi^2 \kappa \simeq 1.18. 
\end{eqnarray}

Form this mass parameter $m_R$ and 
the potential $V^{\rm BPS}(R)$ 
in Eq.~(\ref{eqn:BPS potential data}),  
one obtains a rough estimation of 
the excitation energy of the dilatation mode, 
\begin{eqnarray}
    \omega = \sqrt{\frac{2A^{\rm BPS}}{m_R}} = 0.81\ M_\textrm{KK} \simeq 771\ \textrm{MeV},
    \label{eq:ana_result}
\end{eqnarray}
which seems a larger value than the numerical result 
in Sec.~\ref{sec:dilatation}. 
This estimation can be analytically done 
but has no gravitational effect of $h(w)$ and $k(w)$ 
for both kinetic and potential terms,   
in addition to use of the self-dual BPS instanton. 
On these points, 
the numerical calculation presented 
in Sec.~\ref{sec:dilatation} has been developed.

\end{document}